\newcommand{\pd}[2]{ { \partial {#1} \over \partial {#2} } }

\def\half{\frac{1}{2}}

\def\beqar{\begin{eqnarray}}
\def\eeqar{\end{eqnarray}}

\newcommand{\llabel}[1]{\label{#1}}              

\newcommand{\labeq}[2]{ \begin{equation} \llabel{#1}
{#2}
\end{equation}}

\documentclass[aps,showpacs,twocolumn,prd,preprintnumbers,amsmath,amssymb,letterpaper]{revtex4}

\usepackage{amsmath}
\usepackage{graphics,epsfig,placeins,subfigure,wrapfig}

\begin{document}

\title{\bf Numerical performance of the parabolized ADM (PADM) \\
formulation of General Relativity}

\author{Vasileios Paschalidis${}^1$, Jakob Hansen ${}^1$, and Alexei Khokhlov${}^{1,2}$}

\affiliation{${}^1$ Department of Astronomy and Astrophysics, The
University of Chicago, 5640 S Ellis Ave., Chicago IL 60637  \\
${}^2$ Enrico Fermi Institute, The University of Chicago, 5640 S
Ellis Ave., Chicago, IL 60637 }

\date{\today}

\begin{abstract}

In a recent paper \cite{Paschalidis2} the first coauthor presented a new
parabolic extension
(PADM) of the standard 3+1
Arnowitt, Deser, Misner Êformulation of the equations of
general relativity. By parabolizing first-order ADM in a certain way,
the PADM
formulation Êturns it into a  mixed hyperbolic - second-order parabolic, well-posed
system. The surface of constraints of PADM
becomes a local attractor for all solutions and all possible well-posed
gauge conditions. This paper describes a numerical implementation of PADM and studies its
accuracy and stability in a series of standard numerical tests. Numerical properties of PADM are
compared with those of standard ADM and
its hyperbolic Kidder, Scheel, Teukolsky (KST) extension.
The PADM scheme is numerically stable, convergent and second-order
accurate. The new formulation has better control of the
constraint-violating modes than ADM and KST.
\end{abstract}

\pacs{04.25.Dm, 04.70.Bw}


\maketitle



\section{Introduction}

For years people have tried to obtain analytic solutions of the complex field equations of Einstein's general theory of relativity (GR). Apart from few cases where symmetry is invoked, it is almost impossible to analyze the complicated dynamics in the strong gravitational field regime as described by GR. Approximation methods have been developed over the course of time, but the most promising tool for tackling problems such as gravitational waves arising from binary black hole (BBH) or binary neutron star mergers, gravitational collapse etc., is numerical relativity. 

In the past few years remarkable progress has been made towards achieving
long term and stable evolution of the Einstein equations.  Recently,
particular cases of the the BBH problem were solved \cite{Pret, Camp1, Camp2, Camp3, Baker1, Baker2, Jena}. Despite this remarkable achievement,
the general problem of long term and stable evolution of the GR equations remains open and there is still much work left to be done. There is 
 no theory or prescription to chose what formulation(s) and 
what gauge conditions are suitable for the numerical solution of a given problem. For example, it is well known that the Baumgarte, Shapiro, Shibata, Nakamura (BSSN) formulation \cite{BSSN}, although successful with BBHs, has not been successful with certain spacetimes \cite{testbeds}. In addition to that, there does not seem to exist a definitive explanation of why 
the approaches of \cite{Pret, Camp1, Baker1} perform so well when contrasted to previous efforts. Furthermore, there are astrophysical and theoretical problems of great interest for which the formulations above have not been applied yet and it is not known whether they will prove successful in such cases. Such problems are  the study of the internal structure of black holes and astrophysical phenomena, where except for black holes matter is also involved.

The numerical integration of the Einstein equations is not an easy task because the computations can become unstable and an exponential blow up of the numerical error may occur, even when the formulation employed admits a well-posed initial value problem.  If the numerical techniques employed and gauge and boundary conditions chosen do not suffer from pathologies, perhaps the most important source that can potentially lead to instabilities during a free evolution is the growth of the constraint violating modes.  Over the years several methods have been proposed to deal with this last type of instability \cite{Stark, Abrahams92,Choptuik93,Abrahams93, Abrahams94,Choptuik03a,Choptuik03b,Brodbeck,Gundlach,YonedaShinkai,Holst, Anderson, FiskePRD}.

Of all these approaches, the one which has attracted most attention in recent years takes advantage of the fact that in the ideal case where the constraint equations are satisfied, one has the freedom to add combinations of the constraint equations to the right-hand-side (RHS) of the evolution equations of a given formulation. By virtue of this freedom, it is also possible to introduce in the system of evolution equations terms which act as ``constraint drivers" and  turn the constraint surface into an attractor. This technique is nowadays usually termed as ``constraint damping". 

The goal of many formulations in numerical relativity has been to incorporate such drivers in symmetric or strongly hyperbolic systems without changing the principal part of the evolution equations \cite{Brodbeck, Gundlach}. However, not all formulations with such drivers have been as successful as the generalized harmonic decomposition \cite{Pret}. A possible explanation is that the constraint satisfying modes evolve differently with different but equivalent formulations. Another explanation may be that some formulations may require more efficient damping of the constraint-violating perturbations that are present 
in numerical simulations and lead to all sort of instabilities. 

One way to achieve efficient damping is to construct formulations of GR that under free evolution force all constraint violating modes to evolve according to parabolic equations. Parabolic equations are known for their damping and smoothing properties \cite{PDE_textbook} and this is of extreme importance in numerical relativity. In \cite{Paschalidis2} this goal was achieved by the construction of an evolution system based on the first-order form of the standard Arnowitt, Deser, Misner (ADM) formulation \cite{ADM}, through the addition of appropriate combinations of the derivatives of the constraints and the constraints themselves at the RHS of the ADM evolution equations. We call this evolution system the Parabolized ADM (PADM) formulation throughout this work. It was shown in \cite{Paschalidis2} that the evolution of the constraint equations with the PADM formulation are second-order parabolic, independently of the gauge conditions employed.  This in turn implies that the constraint surface becomes a local attractor. It was finally proved, that the PADM system satisfies the necessary conditions for well-posedness and based on the results of \cite{Paschalidis1} an argument, which indicates strong evidence that the PADM system admits a well-posed initial value problem, was given. 

The purpose of this work is to describe a numerical scheme for solving the PADM equations, test the accuracy and stability of the PADM system, and compare the PADM formulation with the first-order ADM and the Kidder, Scheel, Teukolsky (KST) \cite{KST} formulations. 

The first-order ADM system in conjunction with harmonic or ``$1+\log$" slicing is strongly hyperbolic for the set of one-dimensional solutions considered in this work.  Therefore, we choose the first-order ADM formulation as a basis for comparison. We choose the KST formulation because it is strongly hyperbolic and because the KST and the PADM systems are both extensions of the ADM evolution system, and we want to study how different extensions of the same base system perform numerically. Furthermore,  the only method, to our knowledge, which has been employed to integrate the KST evolution equations is pseudo-spectral methods. This work implements and  demonstrates the numerical performance of the KST formulation with finite difference methods.

The comparison between the three formulations is carried out in a series of standard one- and two-dimensional tests, usually referred to as the ``Apples with Apples" tests or the ``Mexico City" tests \cite{testbeds}. There are four basic tests: a) The evolution of small initial noise on flat spacetime data, b) the evolution of one-dimensional and two-dimensional gauge waves, c) the evolution of a small amplitude 1D and 2D gravitational waves and d) the evolution of polarized Gowdy waves. 

This paper is organized as follows. In section~\ref{Formulation}, we briefly describe the first-order ADM, KST and PADM formulations. In section \ref{code} we describe the numerical scheme we developed. In section \ref{tests} we describe the  standard tests, present the results of the numerical simulations we carried out and compare the numerical performance the aforementioned formulations. In section \ref{code_convergence} we study the convergence of the numerical schemes. We conclude this paper in section \ref{conclusions}.


\section{Formulations \label{Formulation}}


The four-dimensional theory
of general relativity can be cast into a 3+1 decomposition of
spacetime \cite{ADM} by assuming that the spacetime can be foliated by a one-parameter 
family of spacelike hypersurfaces. The spacetime metric is then written in the following form
\labeq{metric}{ds^2=-\alpha^2dt^2+\gamma_{ij}(dx^i+\beta^i dt)(dx^j+\beta^j dt)
}
where $\gamma_{ij}$ is the positive definite 3-metric on the $t=const.$ hypersurfaces, $\alpha$ is the lapse function, $\beta^i$ the shift vector, and 
$x^i$ are the spatial coordinates, $i=1,2,3$.

\subsection{The first-order ADM formulation}

The ADM  formulation and consist 
of two subsets of equations. The first subset is that of the evolution equations, which describe how
the dynamical variables evolve in time. The second subset is that of the constraint equations, which have to be satisfied for
all times. The standard second-order ADM formulation \cite{ADM} has as dynamical variables the 3-metric $\gamma_{ij}$ and the extrinsic curvature $K_{ij}$ of the 3D spacelike hypersurfaces.

The first-order ADM formulation is derived from the second-order one \cite{Paschalidis2} by introducing additional dynamical variables
\labeq{D-defin}{
                      D_{kij} \equiv \partial_k\gamma_{ij},
}
and then deriving the evolution equations for $D_{kij}$. The dynamical variables of the first-order ADM system evolve according to the following equations
\labeq{ADM-gamma}{ \hat\partial_o\gamma_{ij} = -2\alpha K_{ij}, }
\labeq{ADM-K}{
\begin{split}
\hat\partial_o K_{ij}  = & -\nabla_i \nabla_j \alpha+\alpha \left(R_{ij} +  K K_{ij} - 2
\gamma^{mn} K_{im} K_{jn} \right),
\end{split}
}
\labeq{dDijk}{ \hat\partial_o D_{kij}  = -2\alpha
\partial_k K_{ij}-2K_{ij}\partial_k\alpha,}
where $\gamma_{ij}$ is the three-metric, $K_{ij}$ and $K=\gamma^{mn} K_{mn}$ are the extrinsic curvature and its trace respectively, $\nabla_i$ is the covariant derivative operator associated with the three-metric, $R_{ij}$ is the Ricci tensor associated with the three-metric, and $\hat\partial_o=\partial_t-\pounds_\beta$, with $\pounds_\beta$ the Lie derivative along the shift vector $\beta^i$. In equation \eqref{ADM-K}, all partial derivatives of the three-metric have been replaced by $D_{kij}$.

The Lie derivatives of the dynamical variables are 
\labeq{Lie_gamma}{\pounds_{\bf\beta}\gamma_{ij}=\nabla_i\beta_j + \nabla_j \beta_i,}
\labeq{Lie_Kij}{\pounds_{\bf\beta} K_{ij}= (\nabla_i\beta^m) K_{mj} + (\nabla_j\beta^m) K_{mi} + \beta^m \nabla_m K_{ij},}
and
\labeq{Lie_Dkij}{\begin{split}
\pounds_{\bf\beta} D_{kij}=&\ \beta^m\partial_m D_{kij}+D_{mij}\partial_k\beta^m \\
			& \ +2D_{km(i}\partial_{j)}\beta^m+2\gamma_{m(i}\partial_{j)}\partial_k\beta^m.
\end{split}
}

The set of constraint equations is
\labeq{ADM-H}{
               {\cal H}\equiv\quad  R + K^2 - K_{mn} K^{mn} = 0,
}
\labeq{ADM-M}{
              {\cal M}_i\equiv\quad  \nabla_m K^{m}{}_i - \nabla_i K = 0, \quad
              i=1,2,3,
}
\labeq{1stADM-D}{ {\cal
C}_{kij}\equiv\partial_k\gamma_{ij}-D_{kij}=0.}
where $R$ is the trace of the three-Ricci tensor. Just like in the evolution equations, all partial derivatives of the three-metric in \eqref{ADM-H} and \eqref{ADM-M} are replaced by $D_{kij}$. ${\cal H}$ is the Hamiltonian constraint, ${\cal M}_i$ are the momentum constraints and ${\cal C}_{kij}$ are new constraints due to the introduction of $D_{kij}$.  The equations as presented above do not include matter terms, because in this work we focus on vacuum solutions of the Einstein equations. 

At this point we must make an important remark concerning the geometrical interpretation of the ${\cal C}_{kij}$ constraints. The covariant derivative of
any purely spatial tensor $A_{ij}$ is
\labeq{compatibility}{\nabla_k A_{ij}=\partial_k A_{ij}-\Gamma^s{}_{ik} A_{sj}-\Gamma^s{}_{kj} A_{is},
}
where, in the context of the first-order ADM formulation, $\Gamma^k{}_{ij}$ is given by
\labeq{Gamma}{
                  \Gamma^k_{ij} = \frac{1}{2} \gamma^{kn} \left( D_{jin} + D_{ijn} - D_{nij}  \right).
}
 The covariant derivative of the three-metric must be zero, but by virtue of \eqref{compatibility} and \eqref{Gamma} it is straightforward to show
that
\labeq{cov_g}{
\nabla_k\gamma_{ij}={\cal
C}_{kij}.}
Therefore, the geometrical interpretation of the ${\cal C}_{kij}$ constraints is that the covariant derivative is metric-compatible, if and only if
${\cal C}_{kij}=0$.

To close the system, equations
\eqref{ADM-gamma} - \eqref{dDijk} must be supplemented with the
gauge equations for the lapse function and the shift vector. 
Following \cite{Khokhlov}, we write them in a general form as
\labeq{Gauge}{\begin{split} F_a\bigg (      x^b, \alpha, \beta^i,&
\partial_b\alpha,\partial_b\partial_c\alpha, ..., \partial_b\beta^i
,...
      \gamma_{ij}, \partial_b\gamma_{ij},...
\bigg) =0, \\ & a,b,c = 0,...,3, \quad i,j = 1,2,3. \end{split}}

In this work we limit our consideration to algebraic gauges where the lapse function is of the form
\labeq{algebraic_lapse}{\alpha=\alpha(\gamma),}
where $\gamma$ is the determinant of the three-metric $\gamma_{ij}$, and the shift vector $\beta^i$ may be either constant or a fixed
function of the spacetime coordinates ($t,x^i$). In \cite{Paschalidis1,Khokhlov} it was shown that \eqref{algebraic_lapse} makes the evolution
well-posed on the surface of constraints, if $A=\partial\ln\alpha/\partial\ln\gamma>0$. Working with \eqref{algebraic_lapse}, we have the flexibility to use either ``1+log" slicing 
\labeq{1pluslog}{\alpha=1+\ln \gamma}
 or a densitized lapse 
\labeq{dense-lapse}{\alpha=Q \gamma^\sigma,}
where $Q$ is constant or a fixed function of $t,x^i$ and $\sigma$ the densitization parameter.
If \eqref{1pluslog} is used, $A=1/\alpha$. If  \eqref{dense-lapse} is employed, $A=\sigma$.


\subsection{The KST formulation}


The strongly hyperbolic, four-parameter KST modification to the first-order ADM formulation \cite{KST} is
\labeq{KST1}{
\partial_t K_{ij}= (ADM)+\rho \alpha \gamma_{ij}\mathcal{H}+\psi \alpha \gamma^{ab}\mathcal{C}_{a(ij)b}, }
\labeq{KST2}{
\partial_t D_{kij}=(ADM)+\eta \alpha \gamma_{k(i}\mathcal{M}_{j)}+\chi
\alpha \gamma_{ij}\mathcal{M}_k,
}
where $(ADM)$ stands for the RHS of the first-order ADM evolution equations and $\rho, \psi, \eta \mbox{ and } \chi$ are four parameters
of the formulation. In addition to those modifications, Kidder, Scheel and Teukolsky used a densitized lapse gauge \eqref{dense-lapse}.

In this work we use the KST system in conjunction with the more general gauge condition \eqref{algebraic_lapse}. We do so because in \cite{Paschalidis2} it was shown that the KST formulation with \eqref{algebraic_lapse} is strongly hyperbolic, if the KST parameters are 
%
\labeq{c2_c3_1c}{\eta=\frac{6}{5}, \ \ \ \psi=-\frac{5}{6}, \ \ \ \chi=-\frac{2}{5}, \ \ \  \rho\neq 0.
}
It was also shown that for the particular choice of $\rho=-1/3$
%
the number of flat space modes which violate the Hamiltonian constraint is smaller than for any other value of $\rho$.
Thus, the values of the KST parameters that we will use throughout this work are those in \eqref{c2_c3_1c}
 with $\rho=-1/3$.

\subsection{The PADM formulation}

The PADM system is obtained from the first-order ADM formulation by addition of constraints and their derivatives to the RHS of the ADM evolution equations. It  has six parameters and is given by
\labeq{moddgij}{\partial_t \gamma_{ij}=(ADM)+\lambda
\gamma^{ab}\partial_b {\cal C}_{aij},
}
\labeq{moddKij}{\partial_t K_{ij}=(ADM)+\phi \gamma_{ij}\gamma^{ab}\partial_a {\cal M}_b
+\theta \partial_{(i}{\cal M}_{j)},
}
\labeq{moddDkij}{\partial_t D_{kij}=(ADM)+\epsilon \gamma^{ab}\partial_a {\cal C}_{bkij} +\xi\gamma_{ij}\partial_k {\cal H}
+\zeta {\cal C}_{kij}, 
}
where $\lambda, \phi, \theta, \epsilon, \xi, \zeta$ are the six parameters of the formulation. We refer to the added terms of the PADM formulation as the constraint driver. It was shown in \cite{Paschalidis2} that for gauges which do not introduce second-order derivatives of the dynamical variables in the evolution equations the PADM system satisfies the Petrovskii condition for  well-posedness \cite{Gustafsson}, provided that 
\labeq{petrovskii}{\lambda>0, \epsilon >0, \xi<0, \phi<0 ,\theta>0. }
Therefore, in this work we use the PADM equations in conjunction with the algebraic gauge condition \eqref{algebraic_lapse}. 

When the aforementioned conditions are satisfied, the PADM system has the following properties: a) The PADM evolution equations can be classified as a set of mixed hyperbolic  - second-order parabolic quasi-linear partial differential equations. The parabolic character of the equations can be most easily seen by the evolution equation of the 3-metric, whose principal part is given by
\labeq{tmetricPADM}{
\partial_t \gamma_{ij}\simeq\lambda\gamma^{ab}\partial_a\partial_b\gamma_{ij},
}
where $\simeq$ implies equal to the principal part. 

b) The evolution equations of the constraint variables become second-order parabolic partial differential equations (PDEs) independently of the spacetime geometry and the gauge conditions employed. Therefore, the constraint propagation equations admit a well-posed Cauchy problem themselves.  
c) Because of the parabolic structure of the evolution equations of the constraints, the constraint surface becomes a local attractor. All small amplitude, high-frequency constraint-violating perturbations are exponentially damped in time as $\exp(-\lambda\kappa^2 t)$, where $\kappa$ is the magnitude of the wavevector of the perturbations. This in turn implies that the hazardous high frequency perturbations must be damped very efficiently.

In this work we use the PADM formulation with the following set of parameters
\labeq{parabolic_params_val}{\zeta=1, \lambda=\epsilon=\theta=-2\xi=-2\phi=0.02}
 in all our simulations. We do this for two reasons: a) It was shown in \cite{Paschalidis2} that for perturbations about flat space, the Fourier transformed operator of the evolution equations possesses  a complete set of eigenvectors if
\labeq{PADM_condition}{\lambda=\epsilon=-2\xi=-2\phi\neq \theta/2}
and  b) when using explicit numerical schemes to carry out the integration of the PADM evolution equations, the values of the PADM parameters given in \eqref{parabolic_params_val} are small enough to allow for large enough time-steps, while the damping properties of the formulation are still present.

Finally, we note that care must be taken in the choice of the parameters of the PADM system for backwards in time evolutions, e.g, the collapsing Gowdy spacetimes. The backwards in time Cauchy problem for a parabolic equation is ill-posed, because of the existence of exponentially growing modes. The same is true for the PADM formulation if the parameters of the formulation satisfy \eqref{petrovskii}. To overcome this problem we simply have to reverse the signs of the PADM parameters in \eqref{petrovskii}.


\section{Numerical Scheme \label{code}}


We use a finite difference method to carry out computations.  For all formulations we use a staggered spatial mesh in which $\gamma_{ij}$ and $K_{ij}$ are located at the cell centers and $D_{kij}$ are staggered half a grid point in all spatial directions. Moreover, all variables are defined at the same time layer. For ADM and KST we use a third-order Runge-Kutta method to do the integration in time, as this scheme has shown to posses desirable dispersion and dissipation properties compared to other commonly used numerical schemes \cite{Hansen}. We use second-order accurate centered derivative operators to calculate spatial derivatives and a 3rd order accurate parabolic interpolation operator whenever staggered values of the dynamic variables are needed. 

The PADM formulation has both hyperbolic and parabolic terms in the RHS of its evolution equations. 
Since we have implemented an explicit algorithm, if we were to deal with the integration of the PADM evolution equations in the same way as for the KST and ADM systems, we could soon face strict limitations on the speed of the computations because of limitations on the maximum allowed time-step. 
We deal with this limitation by combining the methods of operator split and fractional steps. 

The basic idea of operator splitting  was originally introduced in \cite{Strang}. Here we omit all the details and present the technique in its simplest form.
Let 
\labeq{pde1}{\frac{\partial u}{\partial t} = {\cal L} u}
be a set of PDEs, where $u$ is the column vector of the unknown variables and $\cal L$ is a differential operator. Assume further that
$\cal L$ can be written as a sum of two operators, i.e.,
\labeq{pde2}{{\cal L} u = {\cal L}_1 u + {\cal L}_2 u. }  
Let us now consider the individual equations
\labeq{pde3}{\partial_t u= {\cal L}_1 u, \quad \partial_t u=  {\cal L}_2 u }
and let $C(\Delta t)$ be a finite difference operator for equation \eqref{pde1}, i.e., the solution at time-step $n+1$ is given by $u^{n+1}=C(\Delta t) u^n$. Also,
let $C_1$ and $C_2$ be corresponding finite difference operators for the individual equations of \eqref{pde3}. If the operators $C_1$ and $C_2$ are second-order
accurate and stable for time-step $\Delta t$, then it can be shown that the approximation
\labeq{operator_split}{
C(2\Delta t)=C_1(\Delta t)C_2(\Delta t)C_2(\Delta t)C_1(\Delta t)
}
provides a second-order accurate and stable scheme for \eqref{pde1} for a time-step $2\Delta t$. 

In the discussion above we made no explicit mention to the exact form of the finite difference operators $C_1$ and $C_2$. Therefore, one is free to use any finite difference operators, as long as those are stable and second-order accurate for a time-step $\Delta t$. This means that for any of the operators $C_1$ and $C_2$ we can use the scheme known as fractional steps \cite{Morton}. In this scheme instead of advancing the solution forward a whole time-step $\Delta t$, one advances successively $q$ times, each with a time-step of $\Delta t/q$. Thus, if we choose to do fractional stepping for the $C_2$ operator, we can approximate $C_2(\Delta t)$ as 
\labeq{fractional_steps}{
C_2(\Delta t)=\underbrace{C_2(\Delta t/q)C_2(\Delta t/q) \ldots C_2(\Delta t/q)}_{\mbox{$q$-times}}
}

One may naturally think of the PADM formulation as consisting of two operators which act additively on the dynamical variables, such as those described by equations \eqref{pde1} and \eqref{pde2}. The first operator is that of the RHS of the evolution equations of the ADM formulation and the second one is the constraint driver. There is no unique way to split the entire operator of the RHS of the evolution equations of the PADM formulation, but because of its mathematical structure and for computational efficiency, the most straightforward way to proceed with splitting is to set the ${\cal L}_1$ operator equal to the ADM operator including the low order terms of the constraint driver, and set ${\cal L}_2$ equal to the higher order terms of the constraint driver. We call this ${\cal L}_1$ operator the hyperbolic operator and ${\cal L}_2$ the parabolic operator. If the splitting is done in this way, then the ${\cal L}_2$ operator has smaller number of operations on the dynamical variables compared to any other choice of splitting. This feature is important for computational efficiency as it will be evident shortly. 

For the PADM formulation, we use the techniques described above in the following way. We operator split equations \eqref{moddgij}-\eqref{moddDkij} as is described in the previous paragraph. For the time integration of the hyperbolic terms we use the same technique as that for the ADM and KST formulations. For the time integration of the parabolic terms we use a second-order accurate iterative Crank-Nicolson scheme with two iterations. 
Furthermore, whenever it is required by the numerical stability criterion, we use the fractional steps method to integrate the parabolic terms. We do this by finding the smallest positive integer number $p$, defined by 
\labeq{parabolic_hyperbolic_time_step}{
\Delta t_p = \frac{\Delta t}{  p},}
for which the parabolic time-step $\Delta t_p$ satisfies the stability condition (see discussion on numerical stability below) and then proceed with the integration of the parabolic terms as is dictated by \eqref{fractional_steps}. In equation \eqref{parabolic_hyperbolic_time_step}, $\Delta t$ is the time-step of the hyperbolic terms. The advantage of this algorithm is that even though we may have to use a large $p$ for high resolutions, and thus have to calculate the parabolic terms $p$ times for each hyperbolic time-step, it reduces the computational cost significantly, because the hyperbolic terms have larger computational overhead than the overhead of the parabolic terms. 

An important issue which has to be addressed at this point is the numerical stability criterion of the algorithm for the PADM formulation. Since there is a hyperbolic and a parabolic part in the equations, there arise two stability conditions: a) A condition from the hyperbolic part,  and b) a condition from the parabolic part.  Therefore, we must always make sure that we satisfy both the hyperbolic and the parabolic stability criteria \cite{Morton}. 

The stability criterion for general hyperbolic equations with anisotropic speed of propagation of information is 
\labeq{hyperbolic_stability}{min(\frac{c_x \Delta t}{\Delta x},\frac{c_y \Delta t}{\Delta y},\frac{c_z \Delta t}{\Delta z})=cfl\leq c_1,}
where $c_{x}$, $c_{y}$, and $c_{z}$ are the speeds of propagation of information in the $x-$, $y-$, and $z-$directions respectively, $cfl$ is the Courant-Friedrichs-Levy number, and $c_1$ is a constant which depends on the numerical scheme. For example, for a scalar wave equation with speed unity in conjunction with a 3rd-order Runge-Kutta time integrator and second-order accurate centered spatial derivative operators the constant is $c_1=\sqrt{3/4}$ \cite{Hansen}. 

For most applications in numerical relativity the speed of propagation of information is isotropic and the choice 
\labeq{hyperbolic_stability2}{\frac{\Delta t}{\Delta x}=0.25}
satisfies condition \eqref{hyperbolic_stability}. For this reason in this work we set the time-step for the hyperbolic part of PADM via 
\eqref{hyperbolic_stability2}. We do the same for all computations with the ADM and KST formulations.

Now we turn our attention to the stability criterion of the parabolic part. A Von Neumann analysis of equation \eqref{tmetricPADM} shows that in order to preserve numerical stability the explicit numerical scheme we use must satisfy
\labeq{parabolic_stability}{\lambda \Delta t_p \bigg(\frac{\gamma^{11}}{(\Delta x)^2}+\frac{\gamma^{22}}{(\Delta y)^2}+\frac{\gamma^{33}}{(\Delta z)^2}\bigg)
\le \half.}
Although condition \eqref{parabolic_stability} results from the stability analysis of the principal part of the evolution equation of the 3-metric, all our numerical experiments confirm that if we choose the PADM parameters according to
\labeq{PADM_params}{
\lambda=\epsilon= \theta=-2\xi=-2\phi
}
then if \eqref{parabolic_stability} is satisfied the computations remain stable, and if it is violated the computations soon become unstable.  

Taking all these facts into consideration, we conclude that if condition \eqref{PADM_params} is satisfied, the overall numerical stability criterion of the numerical scheme for the PADM system is 
\labeq{parabolic_stability_c}{ p > \frac{\lambda\big(\gamma^{11}+\gamma^{22}+\gamma^{33}\big)}{2\Delta x},}
with $p=1$ the minimum value $p$ can obtain. In the derivation of the last equation we combined \eqref{parabolic_hyperbolic_time_step}, \eqref{hyperbolic_stability2}, and \eqref{parabolic_stability}.


\section{Numerical Tests \label{tests}}


In this section we briefly describe the standard tests we used and the results of the numerical comparison of the three formulations presented in section \ref{Formulation}. As is usual in the literature, we set up tests which probe both the linear and the non-linear regime of the GR equations.  We closely follow \cite{testbeds} in the setup of the tests, but also modify some of them in order to make the distinction among the three-formulations more apparent. 

For the evaluation of errors we follow \cite{KST_testbeds} and check the accuracy of the integration with a single number given by the following $L_2$ norm 
\labeq{grand_constraint}{
||{\cal C}||_2=\sqrt{\frac{1}{\mbox{Vol}}\int {\cal C}^2\sqrt{\gamma}d^3x},
}
where $\mbox{Vol}=\int \sqrt{\gamma}d^3x$, and 
\labeq{}{
{\cal C}=\sqrt{({\cal H})^2+({\cal M}_i)^2+({\cal C}_{kij})^2}.
}
The squared quantities are equal to the norms of the quantities with respect to the numerical three-metric tensor, e.g., $ ({\cal C}_{kij})^2={\cal C}^{kij}{\cal C}_{kij}$, where all indices are raised with the numerical metric. It is clear from equation \eqref{grand_constraint} that $||{\cal C}||_2=0$, if and only if all the constraints are satisfied. $||{\cal C}||_2$ is called the ``constraint energy."

Similarly, we monitor the accuracy of the numerical solution as compared with the analytic one, by using the  ``error energy" which is given by
\labeq{grand_error}{
||\delta {\cal U}||_2=\sqrt{\frac{1}{\mbox{Vol}}\int \delta{\cal U}^2\sqrt{\gamma}d^3x},
}
where
\labeq{}{
\delta {\cal U}=\sqrt{(\delta\gamma_{ij})^2+(\delta K_{ij})^2 +(\delta D_{kij})^2}, 
}
and where all ``delta" dynamical variables constitute the difference between the numerical and the analytic solution, e.g.
$\delta \gamma_{ij}=\gamma^{\mbox{analytic}}_{ij}-\gamma^{\mbox{numerical}}_{ij}$.  As in the constraint energy, the indices in the error energy are raised with 
the numerical metric. The error energy vanishes, if and only if the numerical solution matches the analytical one.

Just like in \cite{KST_testbeds}, for the Gowdy wave tests we plot the normalized version of those quantities, i.e.,
$||\delta{\cal U}||_2/||{\cal U}||_2$ and $||{\cal C}||_2/||\partial {\cal U}||_2$, where 
\labeq{}{
{\cal U}= \sqrt{(\gamma_{ij})^2+(K_{ij})^2 + (D_{kij})^2}, 
}
and
\labeq{}{
\partial{\cal U}= \sqrt{(\partial_k\gamma_{ij})^2+(\partial_k K_{ij})^2 + (\partial_sD_{kij})^2},}
and where the squared quantities here are with respect to the analytic three-metric tensor. The reason for plotting the normalized errors is that the error depends on the magnitude of the solution.

We set up all tests on a 3D mesh with periodic boundary conditions. We perform all one-dimensional tests on a  cell-centered domain, such that $x\in [0,1]$, with $x_n=(n-\half)\Delta x$, where $n=1,2,\ldots 2^{r}$ and $ \Delta x=\Delta y=\Delta z=2^{-r},$ where  $r \in \mathbb{N}$. In addition, we follow \cite{testbeds2} and set up the domain with four points in the $y-$ and $z-$ directions. In the case of two-dimensional tests the domain is similar. In particular, $x\in [0,1]$, $y\in [0,1]$, $x_n=(n-\half)\Delta x$, $y_n=(n-\half)\Delta y,$ where again $n=1,2,\ldots 2^{r}$, $ \Delta x=\Delta y=\Delta z=2^{-r}$, and have 4 points in the $z$-direction.

\subsection{Robust stability test}

The robust stability test is based on small random perturbations about Minkowski spacetime. The amplitude of the perturbations is sufficiently small so that 
the evolution remains in the linearized regime unless any instabilities arise. The initial data are given by
\labeq{}{
 \gamma_{ij}  =  \delta_{ij}+\epsilon^1_{ij}, \ \ 
K_{ij}  =  \epsilon^2_{ij}, \  \ D_{kij}=\epsilon^3_{kij},
}
where the random numbers $\{\epsilon^1_{ij},\epsilon^2_{ij},\epsilon^3_{kij}\}$ are uniformly distributed in 
$[-10^{-10}/\rho, 10^{-10}/\rho]$, where $\rho=2^{r-4}$. Note that instead of rescaling by $\rho^2$, as is suggested in
\cite{testbeds}, here we rescale by $\rho$. We do this because the constraints for the formulations we are considering
 are first-order, and in order to achieve the same initial violation on the constraints
we have to rescale by $\rho$. We call the above test the ``standard" robust test.

\begin{figure}[b]
\includegraphics[width=0.495\textwidth]{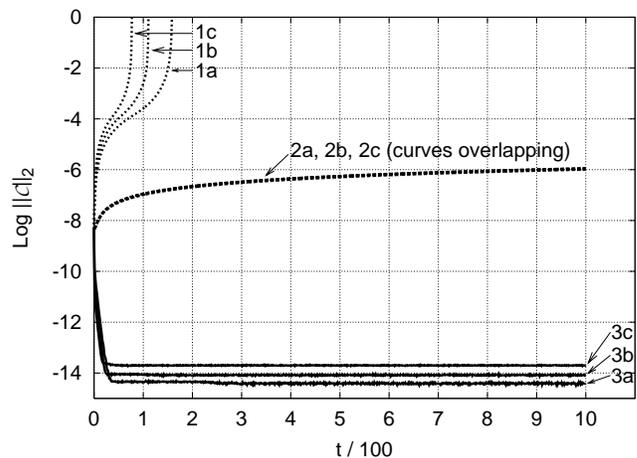}
\caption{\label{fig:1A}  Constraint energies as functions of coordinate time $t$ for the standard robust test with $\alpha=\gamma^{1/2}$. The letters $a, b, c$ correspond to resolutions $r=4, 5, 6$ respectively. Curves $1a$, $1b$ and $1c$ correspond to the ADM formulation, curves $2a$, $2b$ and $2c$ correspond to the KST formulation and curves $3a$, $3b$ and $3c$ correspond to the PADM formulation.}
\end{figure}

Another version of this test requires perturbation by noise not only at the initial time-step, but at every time-step throughout the entire simulation. We call this test the ``strong" robust stability test. We run both versions of the robust test with $r=4,5,6$ for 1000 light crossings or until the code crashes.

\subsubsection{Standard Robust Stability Test \label{strong_robust}}

The results of this test ran with harmonic slicing are summarized in Figure \ref{fig:1A}, where we have plotted the constraint energy for the ADM, KST and PADM formulations as a function of time. The ill-posed nature of the ADM formulation is clearly seen in this figure . The simulations crash fairly quickly and most importantly at earlier times for higher resolutions, as expected for ill-posed systems.

Figure \ref{fig:1A} also shows that the three error energies of the KST runs practically overlap. This is a result of rescaling the amplitude of the random noise with resolution, so that the initial violation of the constraints is of the same magnitude for all resolutions. The KST simulations exhibit noticeable growth in the constraint energy. Close investigation shows that ${\cal C}_{kij}$ grow linearly with time, whereas ${\cal H}$ and ${\cal M}_{i}$ stay roughly constant throughout the entire simulation. Therefore, the growth of the constraint energy comes from ${\cal C}_{kij}$.

The linear growth with time of ${\cal C}_{kij}$  has a very simple explanation. The evolution equations of the ${\cal C}_{kij}$ constraints with the KST formulation are 
\labeq{Ckij}{\hat{\partial}_0 {\cal C}_{kij} = -\eta \alpha \gamma_{k(i}{\cal M}_{j)}-\chi\alpha\gamma_{ij}{\cal M}_k.}
This last equation implies that any small violation of the Momentum constraints will be spilled into  the ${\cal C}_{kij}$ constraints and cause them to evolve with time. Our results of the robust test with the KST formulation show that the RHS of equation \eqref{Ckij} is roughly equal to a non-zero constant. This implies that the ${\cal C}_{kij}$ constraints must grow linearly with time. 

We will now show that the linear growth with time of ${\cal C}_{111}$ with the KST formulation is due to violations of ${\cal M}_x$. We will do this by using our results of the $r=6$ runs of the robust test. Via a least squares fit of the infinity norm of ${\cal C}_{111}$ as a function of time we find that its observed  growth rate is $\partial_t |{\cal C}_{111}|^{observed}_\infty\approx6.5773\cdot 10^{-10}$.

To show that the linear growth of $|{\cal C}_{111}|_\infty$ is due to violations of ${\cal M}_x$ we need to demonstrate that the observed growth rate agrees with the predicted growth rate from equation \eqref{Ckij}. Keeping in mind that the lapse function and $\gamma_{11}$ are practically equal to unity throughout the entire run, and using \eqref{c2_c3_1c} and \eqref{Ckij} we find that the absolute value of the time derivative of ${\cal C}_{111}$ is 
\labeq{dtC111}{
\partial_t |{\cal C}_{111}|\approx\frac{3}{5}|{\cal M}_x|.}
 Using the data from the same simulation we find that the infinity norms of $|{\cal M}_x|_\infty$ stays roughly constant during the entire run and that its time-average is such that $\partial_t |{\cal C}_{111}|^{predicted}_\infty\approx 6.162\cdot 10^{-10}$. The growth rate predicted by  equation \eqref{Ckij} agrees with the observed one within $6.3\%$.  

A consequence of this pathology of the KST system is that given sufficient time or stronger initial perturbations, the linear growth will eventually terminate any KST simulation. 

\begin{figure}[h]
\includegraphics[width=0.495\textwidth]{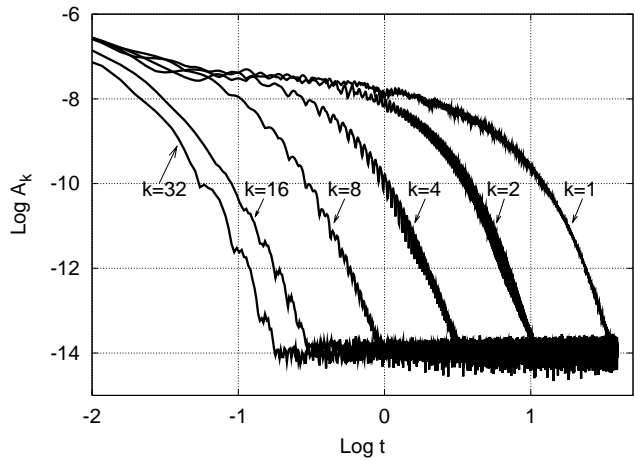}
\caption{Cumulative amplitude $A_k$ (see equation \eqref{Fourier2}) of the Fourier transform of the constraint variables as a function of coordinate time $t$, for harmonics $k=1,2,4,8,16,32$. The plot corresponds to the run of the standard robust test using $\alpha=\gamma^{1/2}$, with the PADM formulation, for resolution $r=6$. 
\label{fig:1C}}
\end{figure}

Figure \ref{fig:1A} shows that the PADM formulation has drastically different behavior than the ADM and KST systems. The constraint energy decreases very rapidly by many orders of magnitude and then reaches a plateau. This formulation damps any small violations of the constraint equations and pushes the evolution back onto the constraint surface. The plateau can be explained as follows. Because of roundoff errors, random noise of the order of $10^{-16}$ is introduced into the system at every time-step. This roundoff error noise causes violations of the constraints of the order of $10^{-16}/dx$ because all our constraints contain first-order derivatives of the dynamical variables. However, $dx$ is roughly of the order of $10^{-2}$ and hence the average magnitude of the constraints when they reach the plateau is expected to lie at roughly $10^{-14}$. This is in accordance with  what we observe if Figure \ref{fig:1A}. 

The plateau occurs at the point where the damping of the constraints essentially balances the constraint violations caused by the roundoff errors.
There is a slight difference in the plateau at which the constraint energy flattens out for different resolutions. It is seen that for the highest resolution $r=6$ (curve 3c), the plateau lies higher than those for resolutions $r=4$ and $r=5$ (curves 3b and 3a respectively). This is because the roundoff errors are on average of the same magnitude and hence the roundoff error perturbations are not scaled as $1/\rho$, as is the case with the initially added random noise. Consequently, the plateau of simulations of the robust test with the PADM system will lie higher with increasing resolutions.

In Figure \ref{fig:1C} we explore the damping properties of the PADM formulation. The plot in the figure was created as follows. Since the simulation is quasi one-dimensional, we focus on a one-dimensional line located along $x=[0,1],y=\Delta y/2,z=\Delta  z/2$. 
Let the constraint variables be denoted by ${\cal C}_i$, where $i=1,2,\ldots, 18$. Along this line and at a given time $t$ we perform a discrete Fourier transform of the $N=2^r$ values of all the constraint variables ${\cal C}_{i,n}, \ n=1,2,\ldots,N$, so that 
\labeq{Fourier1}{
A_{i,k}=\sum_{n=1}^{N} {\cal C}_{i,n} e^{-i\frac{2\pi}{N}k n}, \qquad k=0,1\ldots, N-1
}
where $A_{i,k}$ denotes the fourier harmonic $k$ of the constraint ${\cal C}_i$.
 For a given harmonic, we then sum the amplitudes of all the Fourier pairs of the constraints, i.e., 
\labeq{Fourier2}{
A_k=\sum_{i=1}^{18}|A_{i,k}|.
}
$A_k$ is zero, only if all constraints have zero  contribution to the specific harmonic $k$.

\begin{figure}[t]
\subfigure[{\label{fig:1E}}]{\includegraphics[width=0.495\textwidth]{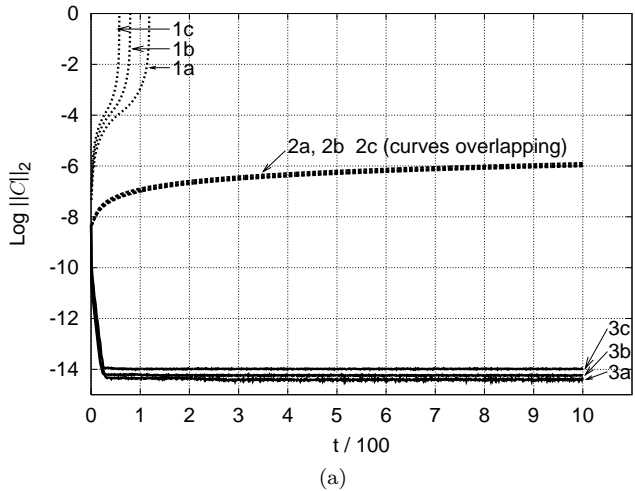}}
\caption{{Constraint energies as functions of coordinate time $t$ for the standard robust test in $1+log$ slicing. The letters $a, b, c$ correspond to resolutions $r=4, 5, 6$ respectively. Curves $1a$, $1b$ and $1c$ correspond to the ADM formulation, curves $2a$, $2b$ and $2c$ correspond to the KST formulation and curves $3a$, $3b$ and $3c$ correspond to the PADM formulation.  \label{fig:1DE} }}
\end{figure}

The quantity $A_k$ as a function of time is what is plotted in Figure \ref{fig:1C} for the standard robust test with the PADM formulation. $A_k$ provides us with a measure which describes the cumulative amplitude of a specific Fourier mode ($k$) over all the constraints. The set of $A_k$ also carries the information of whether any instability arises and on which end of the frequency spectrum it occurs. Figure \ref{fig:1C} shows that the PADM formulation damps the highest frequencies (higher $k$) the most. This is precisely what was predicted in \cite{Paschalidis2} for the PADM system. It is worth noting that the Nyquist frequency ($32$nd harmonic) constraint violations are damped away in less than one light crossing time and all the constraint violating modes are damped away after approximately 25 light crossings.

Finally, we also carried out the standard robust stability test with $1+log$ slicing and with a $\sigma = 1.0$ densitized lapse and we find similar behavior for all formulations. For brevity we only show the results for $1+log$ slicing  (Figure \ref{fig:1DE}). It is again worth noting that the PADM formulation damps the constraints, indicating that the damping properties of the formulation are independent of the gauge choice, as long as the gauge chosen is well-posed according to \cite{Paschalidis1}). This again is in excellent accordance with the predictions of \cite{Paschalidis2}.  

\subsubsection{Strong Robust Stability Test}

This test is much more challenging than the standard robust test and is much closer to reality because in numerical simulations, truncation error will unavoidably be introduced into the solution of the PDE system at every time-step.

\begin{figure}[b]
\includegraphics[width=0.495\textwidth]{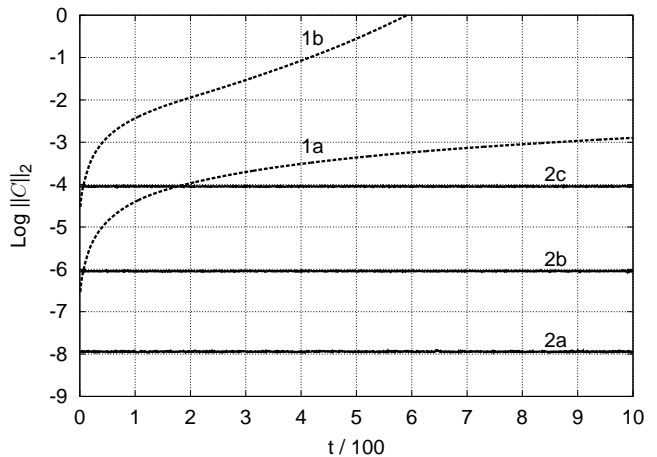}
\caption{Constraint energies as functions of coordinate time $t$ for the strong robust test, with $\alpha=\gamma^{1/2}$ and resolution $r=6$.
The letters $a,b,c$ correspond to noise amplitudes $A=10^{-10}$, $A=10^{-8}$ and $A=10^{-6}$ respectively. Curves $1a$ and $1b$ correspond to the KST formulation. Curves $2a$, $2b$ and $2c$ correspond to the PADM formulation.
 \label{fig:1B}}
\end{figure}

\begin{figure*}
\subfigure[{\label{fig:3A}}]{\includegraphics[width=0.495\textwidth]{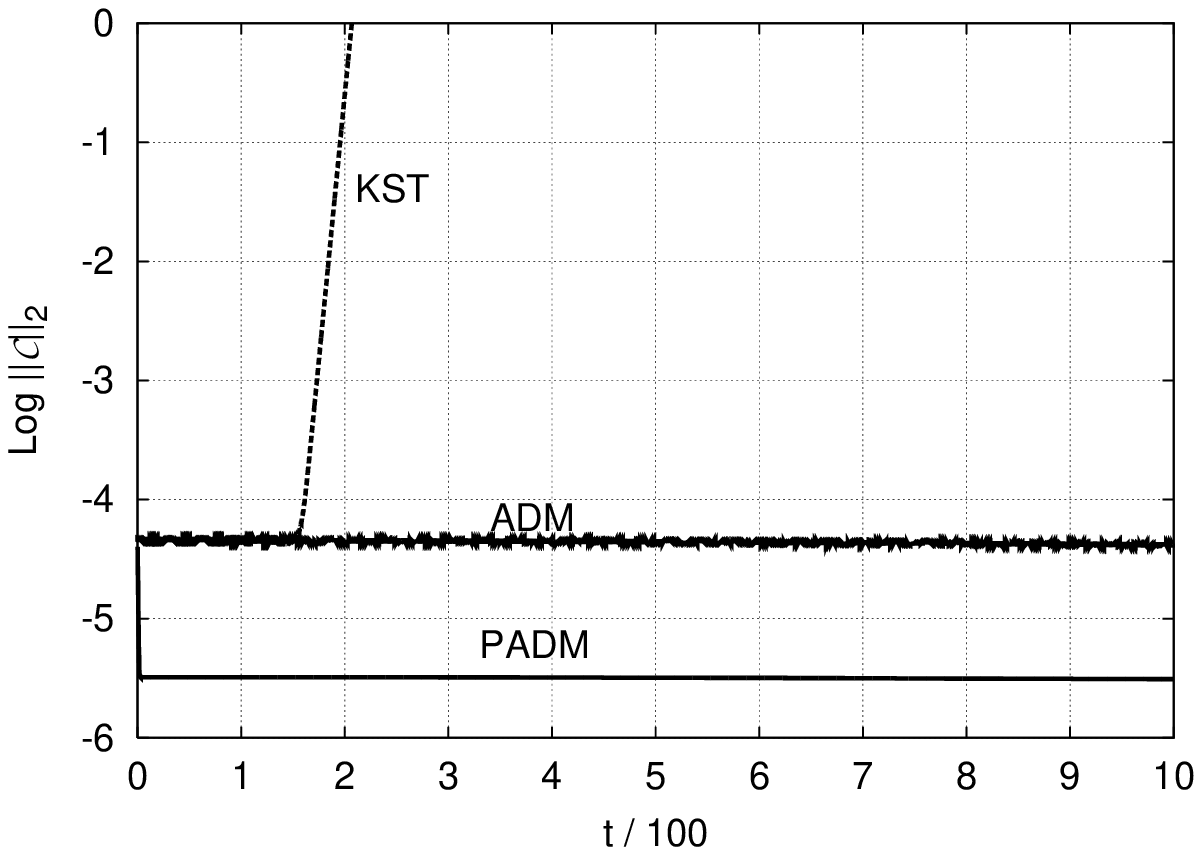}}
\subfigure[{\label{fig:3B}}]{\includegraphics[width=0.495\textwidth]{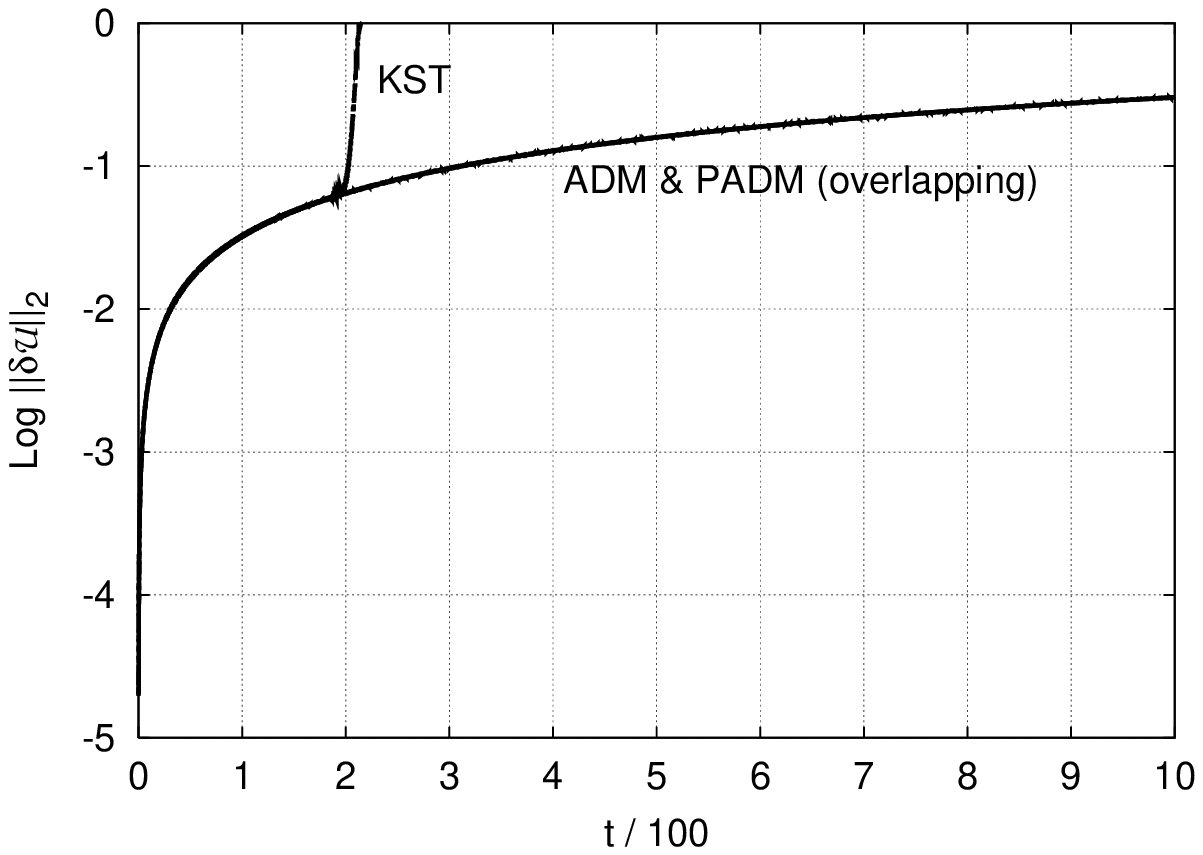}}
\caption{{Errors as functions of coordinate time $t$ for the 1D gauge wave, with $A=0.1$ and resolution $r=7$. On the left panel (a) we show the constraint energies and on the right panel (b) we show the  error energies for the ADM, KST and PADM formulations. The PADM formulation damps the constraints within three light crossing times and reaches the plateau value $||{\cal C}||_2=3.11\cdot 10^{-6}$, while the ADM constraint energy at $t=1000$ is equal to $||{\cal C}||_2=4.15\cdot 10^{-5}$.
 \label{fig:3AB} }}
\end{figure*}

Simulations of this test with the ADM formulation crash almost immediately. For this reason we only show the results for the KST and PADM formulations. In Figure \ref{fig:1B} we  show the constraint energies resulting from the strong robust tests for the KST and PADM systems for various noise amplitudes. 
All plots in this figure correspond to resolution $r=6$. For the KST formulation we see that at amplitude $A=10^{-10}$ (curve $1a$), the constraint energy grows more rapidly than in the standard robust test (Figure \ref{fig:1A}). Nevertheless, the simulation can be run for 1000 light crossing times. However, it is clear that given enough time, the constraint energy will eventually grow so large that the simulation  will be terminated. This can be seen more clearly in curve $1b$, which shows the constraint energy for noise amplitude $A=10^{-8}$. It is evident that in this case the KST runs manage to complete only approximately 600 light crossings before the simulation is so polluted that it crashes.

The results of the PADM runs are dramatically different. Figure \ref{fig:1B} shows no sign of increase of the PADM constraint energy for the first 1000 light crossings. Naturally we do not see the damping behavior to the level shown in Figure \ref{fig:1A}, because strong noise is being pumped into the system at every time-step. However, it is impressive that even with random noise of amplitude $A=10^{-6}$ the constraint energy does not grow throughout 1000 light crossings and the simulations show no signs of pathologies.

\subsection{Gauge wave test \label{gaugewave}}

The metric of the one-dimensional gauge wave test \cite{testbeds} is
\labeq{gauge_metric}{ds^2=-hdt^2+hdx^2+dy^2+dz^2,}
where 
\labeq{pert}{h=1+A\sin\bigg(\frac{2\pi(x-t)}{d}\bigg),}
where $A$ is the amplitude of the gauge wave and $d$ the size of the evolution domain. 
For the first-order ADM formulation \eqref{gauge_metric}, \eqref{pert} is equivalent to
\labeq{1D_gauge_sol}{\begin{split}
\gamma_{11}= & 1+A\sin\bigg(\frac{2\pi (x-t)}{d}\bigg),\\
K_{11}=& -\frac{\pi A}{d\gamma_{11}}\cos\bigg(\frac{2\pi(x-t)}{d}\bigg) \\
D_{111}=& \frac{2\pi A}{d}\cos\bigg(\frac{2\pi (x-t)}{d}\bigg),\\
\end{split},
}
$\gamma_{22}  =\gamma_{33} =1$, and all other dynamical variables are zero.
A further coordinate transformation 
\labeq{2D_transf}{
x=\frac{1}{\sqrt{2}}(x'-y'),\ \ y=\frac{1}{\sqrt{2}}(x'+y'),
}
yields the two-dimensional diagonally traveling gauge wave.

The gauge wave stability test has proven to be one of the most challenging tests for most evolution schemes mainly due to 
 the existence of a one-parameter family of exponential, harmonic gauge solutions \cite{shifted_gauge_wave} which correspond to the metric
\labeq{gauge_metric_exp}{ds^2=e^{\mu t}h(-dt^2+dx^2)+dy^2+dz^2,}
for arbitrary $\mu$. The gauge wave test \eqref{gauge_metric} corresponds to $\mu=0$, but in any numerical implementation of harmonic gauge conditions the exponential modes, $\mu\neq 0$, are likely to be excited by truncation error. This eventually causes the numerical solution to veer off the analytic one.

The growth of these exponential gauge modes can be avoided by the incorporation of a set of semi-discrete conservation laws into the principal part of the evolution equations of a  harmonic formulation \cite{shifted_gauge_wave}. However, these conservation laws are specific to the gauge-wave solution and it is not known whether similar conservation laws exist for arbitrary spacetimes.

The lapse function for the gauge wave test is harmonic and we use $\alpha=\gamma^{1/2}$ in our simulations. Since these coordinates are harmonic and we do not incorporate any conservation laws in our scheme, we do expect to excite the exponential gauge modes. We run  both  the 1D and the 2D gauge wave tests with amplitudes $A=0.01,0.1$ and resolutions $r=5,6,7$. The  1D gauge wave is run for 1000 light crossings or until the code crashes and  the 2D gauge wave is run for 100 light crossings or until the code crashes. Below we describe the results of this test for each formulation separately.

\subsubsection{The low amplitude 1D gauge wave}

The $A=0.01$ 1D gauge wave presented no apparent difficulty for any of the formulations used in this work. For brevity, we only describe the results of this test because the high amplitude gauge wave, discussed directly below, is much more interesting. The error energies as functions of time, with all three formulations, completely overlap. However, while the KST and ADM constraint energies overlap, the PADM constraint energy lies at least two orders of magnitude below those two. The PADM formulation damps the constraints very fast and reaches the plateau value $||{\cal C}||_2=3.17 \cdot 10^{-8}$ within three light crossing times.

\subsubsection{ The high amplitude 1D gauge wave}

Our results for the high amplitude 1D gauge wave are summarized in Figure~\ref{fig:3AB}, where we show the error and constraint energies as functions of the coordinate time $t$ for resolution $r=7$.

\paragraph{The behavior of the ADM formulation:}

Figure \ref{fig:3A} shows that the constraint energy in the high amplitude gauge wave runs with the ADM formulation does not grow. However, Figure \ref{fig:3B} shows that the ADM error energy grows quickly with time and overlaps with the PADM error energy. For both systems, the source of this growth is the excitation of exponential harmonic gauge modes. We do not study here this subject because we address it thoroughly below in the context of the PADM results of the 1D gauge wave test.

\paragraph{ The behavior of the KST formulation:}

Figure \ref{fig:3AB} show that the KST system cannot pass the high amplitude gauge wave test because the computations do not complete 1000 light crossing times. The KST constraint energy suddenly begins to grow exponentially at $t\approx 160$, and later on at $t\approx 200$ the KST error energy blows up.

\begin{figure}[h]
\includegraphics[width=0.495\textwidth]{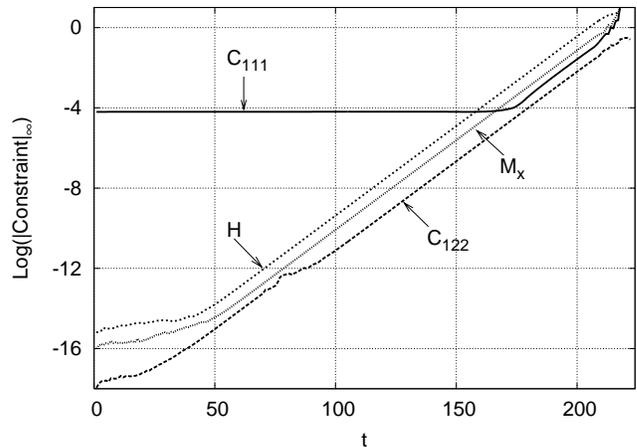}
\caption{Infinity norms of the ${\cal C}_{111}, {\cal H}$, ${\cal M}_x$ and ${\cal C}_{122}$ constraints as functions of the coordinate time $t$. The plot corresponds to the KST run for the gauge wave test, with $A=0.1$ and resolution $r=7$.
\label{fig:3CC}}
\end{figure}

We will now show that the unexpected behavior of the KST system in the simulations of the $A=0.1$ gauge wave is due to excitation of unwanted degrees of freedom by violations of the momentum constraints. If there is no noise introduced in an evolution system, the only evolving dynamical variables for the  one dimensional gauge wave must be $\gamma_{11}, K_{11}$, and $D_{111}$. This implies that the only constraint that evolves is ${\cal C}_{111}$. Furthermore, ${\cal H}$ and ${\cal M}_x$ must evolve due to roundoff error, but must be of the order of $10^{-16}$. However, this is not the case with the KST system. It is straightforward to show by use of \eqref{KST2} that the KST equations excite $D_{122}$, $D_{133}$, $D_{212}$, and $D_{313}$ because ${\cal M}_x$ is not exactly satisfied. These dynamical variables  in turn excite $\gamma_{22}, \gamma_{33}, \gamma_{12}, \gamma_{13}$, $K_{22}, K_{33}, K_{12}, K_{13}$ and eventually all the constraints these variables are involved in, i.e., ${\cal C}_{122}$, ${\cal C}_{133}$, ${\cal C}_{112}$, ${\cal C}_{113}$, ${\cal M}_{i}$, and ${\cal H}$, grow rapidly with time. The constraint violations soon become very large that a non-linear instability develops and the evolution is terminated. 

The excitation of unwanted degrees of freedom can be seen in Figure~\ref{fig:3CC}, where we show the infinity norms of ${\cal C}_{111}$, ${\cal C}_{122}$, ${\cal M}_x$ and ${\cal H}$ as functions of the coordinate time $t$. An instability causes ${\cal C}_{122}$, ${\cal M}_x$ and ${\cal H}$ to grow exponentially  in time from very early on in the evolution. When the violations of ${\cal C}_{122}$, ${\cal M}_x$ and ${\cal H}$ grow large enough, the solution  gets spoiled and at $t\simeq 200$, the numerical solution is so polluted that the simulation crashes.

To prove that that the instability of the evolutions of the gauge wave test with the KST formulation is caused by the excitation of $D_{122}$, $D_{133}$, $D_{212}$ and $D_{313}$ by roundoff error in ${\cal M}_x$, we also ran a ``partially constrained" evolution of the gauge wave with the KST system, where we set the $D_{122}$, $D_{133}$, $D_{212}$ and $D_{313}$ equal to the exact values these should have at every time-step, i.e., $0$. The result of these runs was that the KST formulation can evolve the gauge wave for 1000 light crossing times without any problems. Furthermore, we find that the KST error and constraint energies completely overlap with those of the ADM formulation. We do not show plots of these simulations, but any  plots can be made available by the authors upon request.

These results indicate that simulations of the gauge wave with formulations which add multiples of the constraints to the RHS of evolution equations are likely to be terminated by the same type of instability that terminates corresponding simulations with the KST formulation. An example of such formulation is the BSSN system and it is known that until now no BSSN based numerical scheme has been able to give satisfactory gauge wave simulations with $A=0.1$ \cite{testbeds2,BSSN_testbeds,Kiuchi}. Moreover, the hyperbolic system studied in \cite{AA_testbeds} demonstrated similar behavior. We do not analyze this subject any further, as it is beyond the scope of this paper and will be addressed in a future work.

\paragraph{The behavior of the PADM formulation:\label{PADM_gauge}}
Figure \ref{fig:3A} shows that the PADM constraint energy does not grow with time and that it is over an order of magnitude less than the corresponding ADM constraint energy. However, Figure \ref{fig:3B} shows that the PADM error energy grows quickly in time. We will now show that the source of this growth in the gauge wave simulations is the excitation of exponential harmonic gauge modes \eqref{gauge_metric_exp}.

To obtain direct evidence that exponential harmonic gauge modes are excited during a simulation we must show that at any given time $t$ 
\labeq{normalization}{\gamma_{xx}=f(t)\gamma_{11},} 
where $\gamma_{xx}$ is the numerical solution corresponding to $\gamma_{11}$ of equation \eqref{1D_gauge_sol}, and where $f(t)$ is a normalization factor to be determined. We do not assume that $f(t)=e^{\mu t}$ with fixed $\mu$ because truncation errors are continuously introduced into the numerical solution and hence different values of $\mu$ are likely to be excited as the evolution proceeds. 

We will now show how one can calculate $f(t)$ numerically. The spatial average of  $\gamma_{11}$ is unity, because $\int^1_0\sin(2\pi(x-t)/d)dx=0$. Therefore, for an exponential harmonic gauge mode, $\gamma_{11}({\mu, x,t})=e^{\mu t}\gamma_{11}(x,t)$, the analytical counterpart $f_a(t)$ of $f(t)$ is 
\labeq{ft}{f_a(t)=\int_0^1\gamma_{11}(\mu,x,t) dx.}
The result of the integration of equation \eqref{ft} is $f_a(t)=e^{\mu t}$ as expected. To find $f(t)$ all we must do is to replace $\gamma_{11}(\mu,x,t)$ in \eqref{ft} by $\gamma_{xx}$. $f(t)$ must be different for different resolutions, because $\gamma_{xx}$ changes with resolution. For this reason we denote the normalization factor of \eqref{normalization} at the $n-$th time-step by $f^n_r$ and we calculate it as
\labeq{}{
f^n_r=\frac{1}{2^r}\sum_{i=1}^{2^r}\gamma^n_{xx,i}
}
where $\gamma^n_{xx,i}$ denotes the value of the numerical solution at the $i$-th grid point and time $t_n$. 

\begin{figure}[b]
\includegraphics[width=0.495\textwidth]{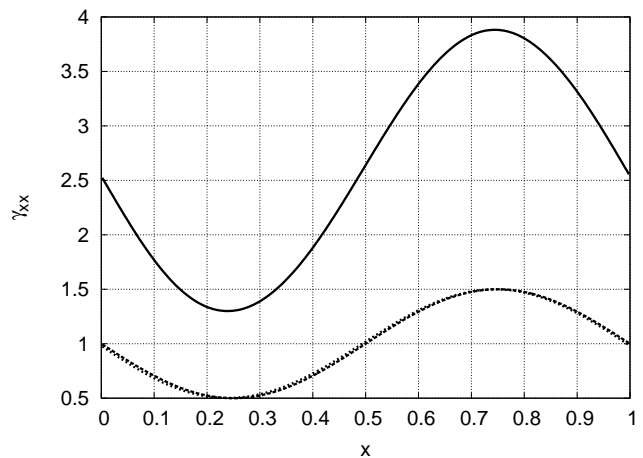}
\caption{Snapshot of the solution of the $\gamma_{xx}$ component for the 1D gauge wave, with $A=0.5$, resolution $r=8$ and $t=300$. 
The continuous curve corresponds to the PADM numerical solution, while the dashed curve is the exact solution. The dotted curve corresponds to the normalized
PADM numerical solution, which is essentially overlapping with the exact solution.
\label{fig:3DD}}
\end{figure}

\begin{figure}[h]
\includegraphics[width=0.495\textwidth]{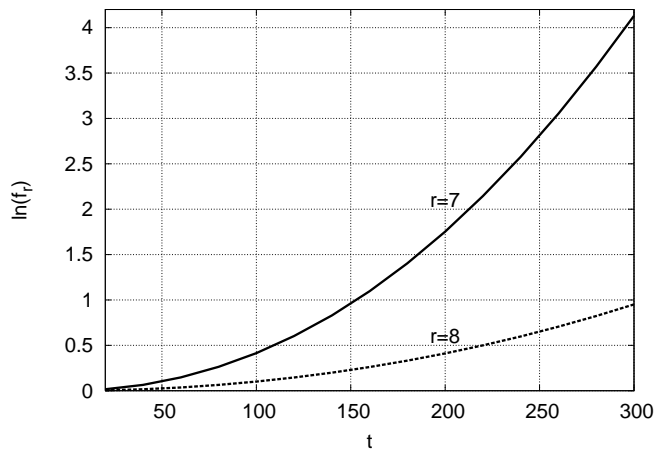}
\caption{Normalization factor $f^n_r$ as a function of time. The plot corresponds to the PADM simulations of the 1D gauge wave, with amplitude $A=0.5$ and resolutions $r=7,8$. 
\label{fig:3G}}
\end{figure}

\begin{figure*}
\subfigure[{\label{fig:3C}}]{\includegraphics[width=0.495\textwidth]{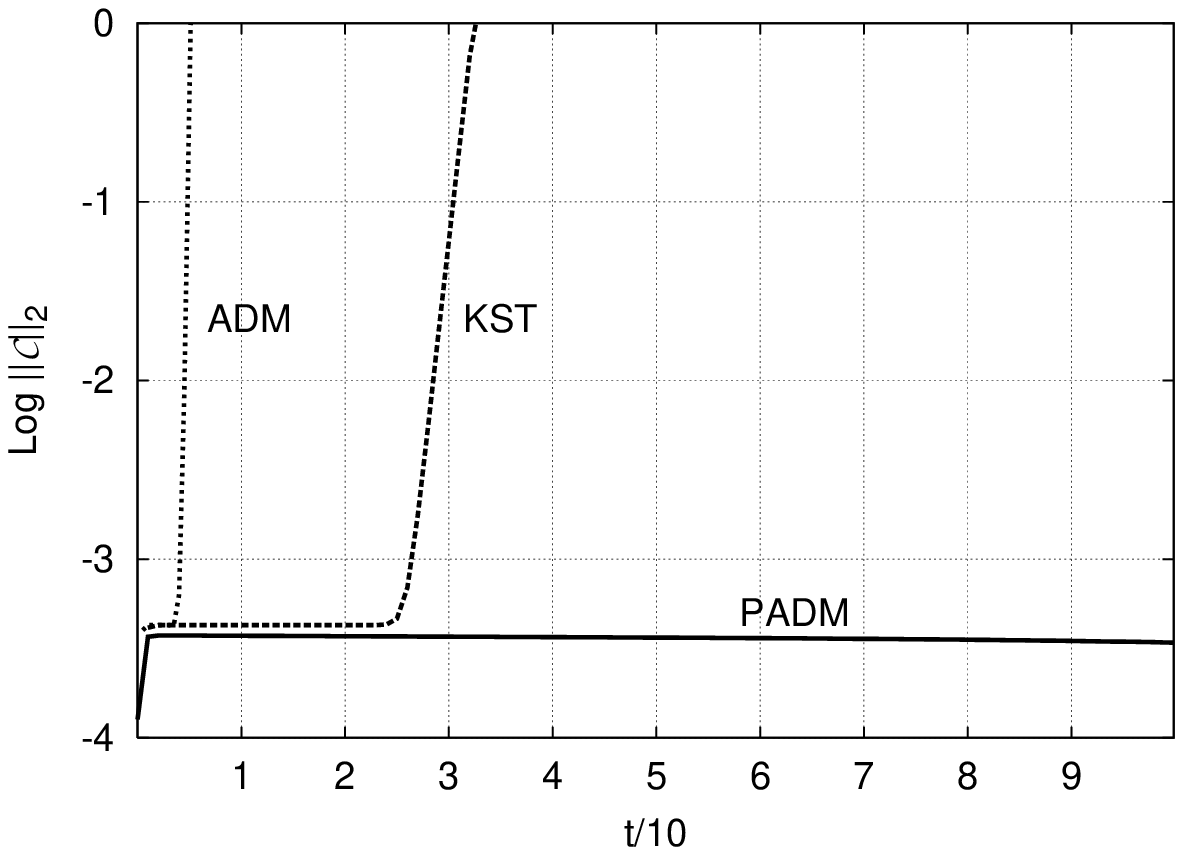}}
\subfigure[{\label{fig:3D}}]{\includegraphics[width=0.495\textwidth]{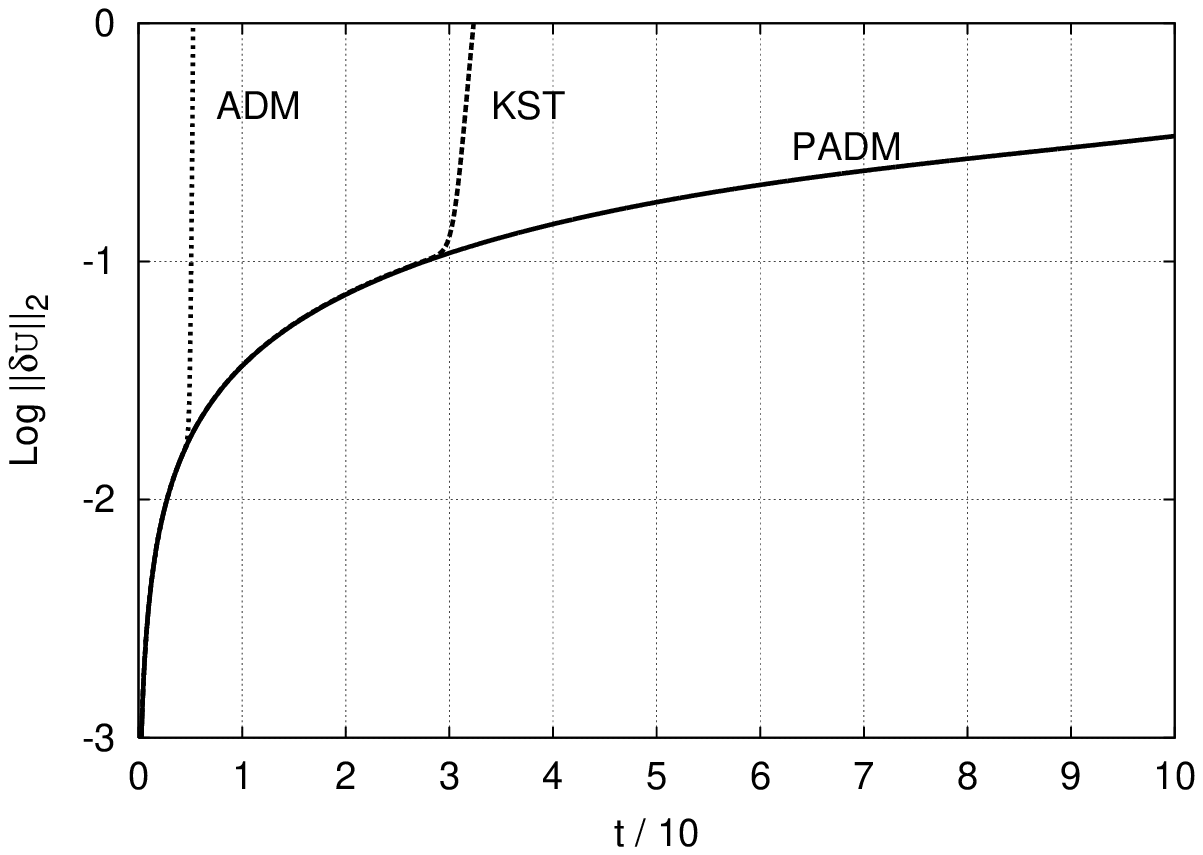}}
\caption{{Errors as functions of coordinate time $t$ in units of light crossing times for the strong two-dimensional gauge wave and resolution r=7. On the left panel (a) we show the constraint energies and on the right panel (b) we show the error energies for the ADM, KST and PADM formulations.  \label{fig:3CD} }}
\end{figure*}

To demonstrate that equation \eqref{normalization} holds true  in our simulations of the gauge wave, we use the data of the gauge wave runs with $A=0.5$. 
In Figure \ref{fig:3DD} we show a snapshot of $\gamma_{xx}$, $\gamma_{xx}/f^n_r$ and $\gamma_{11}$ at $t=300$. The continuous curve corresponds to the numerical solution with the PADM formulation and the dashed curve to the analytical one. The rescaled numerical solution corresponds to the dotted curve, but is essentially overlapping with the exact solution. We find that the rescaled numerical solution overlaps with the analytic one not only at $t=300$, but also at all times.
This proves that \eqref{normalization} holds true in our simulations and in turn provides us with direct evidence that the main source of the growth of the error is the excitation of exponential harmonic gauge modes by truncation error. The same analysis as presented above shows that the same source of error causes the growth of the PADM error energy seen in Figure \ref{fig:3B}. 

We now turn our attention to the properties of function $f(t)$ and we will show that in our simulations $f(t)=e^{a t^2}, \ a >0$. Figure \ref{fig:3G} shows the natural logarithm of $f^n_r$ as a function of time for the gauge wave runs with $A=0.5$ and resolutions $r=7,8$. If truncation errors excited a given $\mu$ of the solution \eqref{gauge_metric_exp}, the curve $t-\ln f^n_r$ should be a straight line with the slope equal to $\mu$. However, the growing slope of the curves in Figure \ref{fig:3G} indicates that the normalization factor grows faster than exponentially. If we interpret the local in time slope of these curves as the value of $\mu$, the former implies that $\mu$ grows as time goes by. 

In order to find how $\mu$ evolves with time we assume that $f(t)=e^{\mu t}$ with time-varying $\mu$. If this is the case, $\mu=\ln f(t)/t$. Thus, to find $\mu(t)$ we need to study how $\ln f^n_r/t$ evolves with time. We do not show $t-\ln f^n_r/t$ plots, but our results indicate that $\ln f^n_r/t$ grows linearly with time. This implies that $\mu(t)=a\cdot t$ and thus $f(t)=e^{at^2}$. For example, by using the $r=8$ data we find via  a least squares fit that $a=1.05915\cdot10^{-5}$. As the resolution increases, $a$ decreases. This is also clear  in Figure \ref{fig:3G}, because  at any given time $f^n_8<f^n_7$.
This is consistent with the fact that with increasing resolution the truncation errors become smaller and hence the solution approaches $\mu=0$. 

The existence of exponential harmonic gauge modes is known to be a property of harmonic coordinates only. Thus, we anticipate that simulations of the gauge wave would not demonstrate the same error growth in other gauges. This analysis will the subject of future work and for now we conclude that the performance of the PADM formulation in the 1D gauge wave test is better than those of the ADM and KST formulations, since the overall PADM error is less than the overall ADM error and much less than overall KST error.

\subsubsection{The low amplitude 2D gauge wave}

The simulations of the 2D gauge wave with $A=0.01$ are not as interesting as those of the 2D strong gauge wave. For this reason we only describe the results of the weak 2D gauge wave here. The PADM and KST formulations have no problem  completing 100 light crossing times without any pathologies in the constraint error. The constraint energies of the  simulations with PADM are damped and lie lower than the simulations with KST. On the other hand, both the PADM and KST systems experience growth in the error energy due to the excitation of exponential harmonic gauge modes.  In contrast to the KST and PADM results, the ADM formulation cannot evolve the initial data for more than 66 light crossing times. 

\subsubsection{The high amplitude 2D gauge wave}

Our results of the strong two-dimensional gauge wave are summarized in Figure \ref{fig:3CD}, where we show the error and constraint energies for $r=7$ for the three evolution systems considered here. 

The results with the ADM formulation are drastically different than the corresponding results in the 1D case. In the gauge condition we employed and in 1D the ADM evolution equations can be shown to be strongly hyperbolic, but in 2D they are only weakly hyperbolic. Therefore, due to the ill-posed nature of the ADM formulation the simulations of the 2D gauge wave crash quickly, and most importantly they crash faster with increasing resolution. We find that for $r=5,6,7$ the ADM runs crash after $16,9,5$ light crossing times respectively. 

The KST evolution equations are strongly hyperbolic, and hence the KST runs of the 2D gauge wave are longer than those of the ADM system. However, the simulations with the KST system crash very early on and cannot complete more than 30 crossing times.  We have not been able to understand the mechanism via which the KST formulation crashes in this test. Similarly to the 1D case, there are $C_{3ij}$ constraints which are excited because of violation the momentum constraints, and then grow exponentially with time . However, when the computations stop the values of $C_{3ij}$ are of the order of $10^{-10}$ and this level of violation is not strong enough to explain the unexpected behavior of the KST system. The violations of the momentum constraints introduce significant error into the $C_{1ij}$ and $C_{2ij}$, but all constraints seem to grow exponentially at the same time, and thus we have not been able to attribute the blow up of the simulations to same source as in the 1D gauge wave. Finally, we note that the instability which causes the evolutions to crash occurs at a time-scale which is  resolution-independent.

The results of the gauge wave simulations with the PADM formulation are drastically different. Figures \ref{fig:3CD} show that PADM successfully completes $100$ light crossing times. In addition to that and as expected, the PADM formulation has better control of the constraints than the ADM and KST formulations, see Figure \ref{fig:3C}. Figure \ref{fig:3D} shows that the PADM error energy experiences rapid growth.  The same is true for the ADM and PADM error energies until the ADM and KST runs crash. This growth results because of  excitation of exponential harmonic gauge modes as in the 1D gauge wave.

Finally, we  note that for resolutions $r=5$ and $r=6$ the PADM system manages to evolve the 2D strong gauge wave initial data for $56$ and $98$ light crossing times respectively. We have not been able to understand the mechanism via which the simulations of the 2D gauge wave with the PADM system crash for low resolutions, but it seems to be correlated with the exponential growth of the solution. Nevertheless, the indisputable conclusion of the gauge wave simulations is that the numerical performance of the PADM system is much better than those of the KST and ADM formulations.

\subsection{The linear wave test}

The linear wave test is a traveling gravitational wave of amplitude small enough so the evolution remains in the linear regime. The metric is given by
\labeq{linear_metric}{
ds^2=-dt^2+dx^2+(1+b)dy^2+(1-b)dz^2,
}
where \labeq{amp}{b=A\sin\bigg(\frac{2\pi (x-t)}{d}\bigg),}
 where $A$ is the amplitude of the wave and $d$ the size of the evolution domain. The suggested amplitude for this test is $A=10^{-8}$ \cite{testbeds}.
For the first-order ADM formulation \eqref{linear_metric}, \eqref{amp} is equivalent to
\labeq{}{
\gamma_{11}=1, \ \gamma_{22}=1+b, \  \gamma_{33}=1-b,
}
\labeq{}{
K_{22}=-K_{33}=-\frac{\pi A}{d}\cos\bigg(\frac{2\pi (x-t)}{d}\bigg),
}
\labeq{}{
D_{122}=-D_{133}=\frac{2\pi A}{d}\cos\bigg(\frac{2\pi (x-t)}{d}\bigg),
}
and all other dynamical variables are zero.

Equation \eqref{linear_metric} suggests geodesic slicing  for the lapse function, but the test can be run both with 
$\alpha=\gamma^{1/2}$ and $\alpha=1+\ln\gamma$. This is so because for \eqref{linear_metric} 
$\gamma=1-b^2$. However, $b$ is so small that  $\alpha=\gamma^{1/2}\approx 1-b^2/2\approx 1-5\cdot 10^{-17}\approx 1$. Thus, because of roundoff error a computer cannot ``see" the difference between geodesic slicing and harmonic slicing. Similarly, it can be shown that ``1+log" slicing is compatible with \eqref{linear_metric}. As in the gauge wave case, the 2D linear wave can be obtained from \eqref{linear_metric} by the coordinate transformation \eqref{2D_transf}.

\begin{figure}[h]
\includegraphics[width=0.495\textwidth]{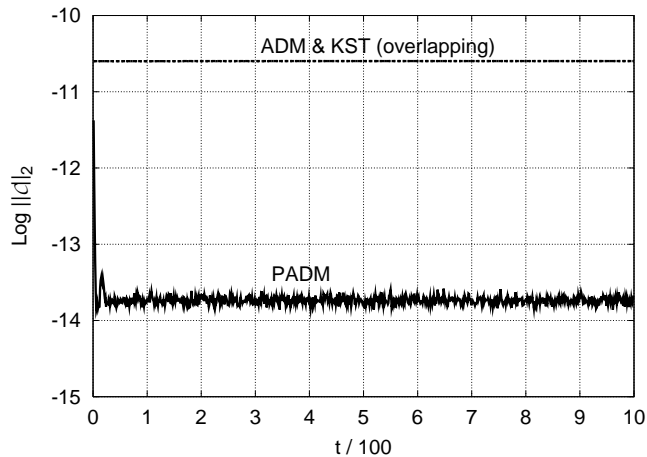}
\caption{Constraint energies as functions of coordinate time $t$ for the 1D linear wave for resolution $r=6$. 
\label{fig:2A}}
\end{figure}

\begin{figure}
\includegraphics[width=0.495\textwidth]{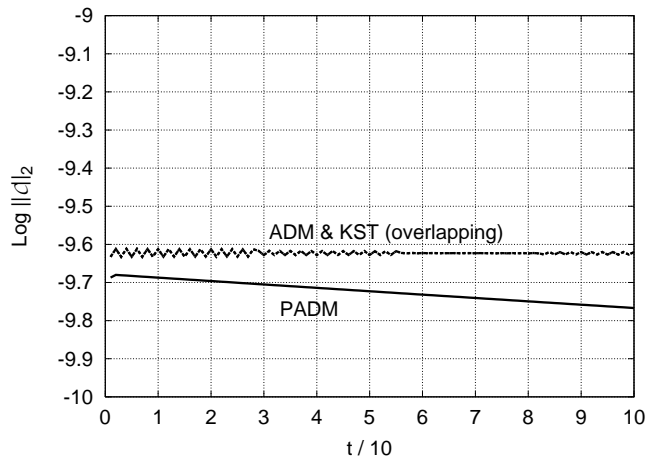}
\caption{Constraint energies as functions of coordinate time $t$ for the 2D linear wave test for resolution $r=6$.
 \label{fig:2B}}
\end{figure}

We carry out simulations of both a one-dimensional and a two-dimensional wave
with harmonic slicing $\alpha=\gamma^{1/2}$ and resolutions $r=5,6,7$. The simulation time is 1000 light crossings in the 1D case and 100 light crossings in the 2D case. The results of our runs are summarized in Figures~\ref{fig:2A} and~\ref{fig:2B} where we show the constraint energies for the three formulations in the 1D and 2D version of the test respectively. For brevity we do not show the error energies, because all formulations can easily handle this test and the essential difference between them lies in the behavior of the constraint energies.

In the 1D simulations (Figure~\ref{fig:2A})  the constraint energies of the KST and ADM formulations overlap and stay constant throughout the 1000 light crossings of the simulation. On the other hand, the PADM formulation pushes the constraint energy down close to the roundoff error level. There is a difference of at least 3 orders of magnitude between the constraint energy of the PADM system and those of the KST and ADM systems. 

In Figure~\ref{fig:2B} we show the constraint energies of the three formulations considered in this work for the 2D linear wave test. In the case of the PADM formulation we output the data at every light crossing time. However, outputting the constraint energies for the KST and ADM formulations every light crossing time, results in a false picture of the behavior of the KST and ADM formulations for this test, as the constraint energies can be seen to grow in time. What really happens is that the constraint energies wildly oscillate roughly around the value $2.4\cdot 10^{-10}$ and they show no sign of growth in time. For this reason, instead of outputting the constraint energy at every light crossing time, we output at every time-step and average the constraint energy through each light crossing. Then this average value is assigned to be the value of the constraint energy at every light crossing time, which is what we have plotted in Figure~\ref{fig:2B} for the KST and ADM systems.

From this figure it is clear that the ADM and KST constraint energies stay roughly constant throughout the entire run. For the 2D high amplitude gauge wave the damping properties of the PADM formulation are still evident, but not at the same level as in the 1D case. Nevertheless, the PADM system controls the constraint violations better than the KST and ADM systems.

\subsection{Gowdy Spacetimes}

The polarized Gowdy T3 spacetimes are solutions of the Einstein equations which describe an expanding (or contracting) universe containing plane polarized gravitational waves, see for example \cite{testbeds} and references therein.

\subsubsection{Expanding solution \label{exp_gowdy}}

The expanding Gowdy metric is usually written as

\labeq{gowdymetric1}{ds^2=t^{-1/2}e^{\lambda/2}(-dt^2+dz^2)+tdw^2,
}
where
\labeq{gowdymetric2}{ 
dw^2=e^Pdx^2+e^{-P}dy^2.
}

The solution which is considered standard is the one with
\labeq{lambda_zt}{\begin{split}
\lambda(z,t)  = & 2 \pi ^2t^2 \left(J_0(2 \pi  t)^2+J_1(2 \pi  t)^2\right) \\
		 & -2 \pi t J_0(2 \pi t) J_1(2 \pi  t) \cos ^2(2 \pi  z)   \\ 
			 & + \pi  J_0(2 \pi )   J_1(2 \pi )-2 \pi ^2 \left(J_0(2 \pi )^2+J_1(2 \pi )^2\right)
\end{split}
}
and 
\labeq{P_zt}{ 
P(z,t)=J_0(2 \pi  t) \cos (2 \pi  z),
}
where $J_n$ denotes the Bessel function of the first kind of order $n$. 

The non-zero dynamical variables in the context of the first-order ADM formulation can be directly derived from \eqref{gowdymetric1} and they are presented in appendix \ref{gowdy_exp_sol}.
By use of equations \eqref{gowdymetric1} and \eqref{gowdymetric2}, it is straightforward to show that the lapse is harmonic and that
\labeq{expanding_lapse}{
\alpha=t^{-1}\gamma^{1/2}.
}

The expanding Gowdy wave test has proved to be one of the most challenging problems in numerical relativity. This is because  
the metric components grow exponentially with time and hence the truncation errors introduced in the numerical solution grow with time, if the resolution is fixed. Consequently, the numerical solution soon veers off the analytical one. Another result of the exponential growth is that very soon the dynamical variables grow so large that the computations cannot be handled by using standard 64-bit 
floating-point arithmetic. Therefore, all simulations of the expanding Gowdy solution are expected to be terminated \cite{testbeds}.

We start our expanding gowdy wave simulations at $t=1$ and evolve the initial data forward in time using the densitized lapse \eqref{expanding_lapse}.
We run the test for resolutions $r=6,7,8$ and until the code crashes.

\begin{figure*}
\subfigure[{\label{fig:4C}}]{\includegraphics[width=0.495\textwidth]{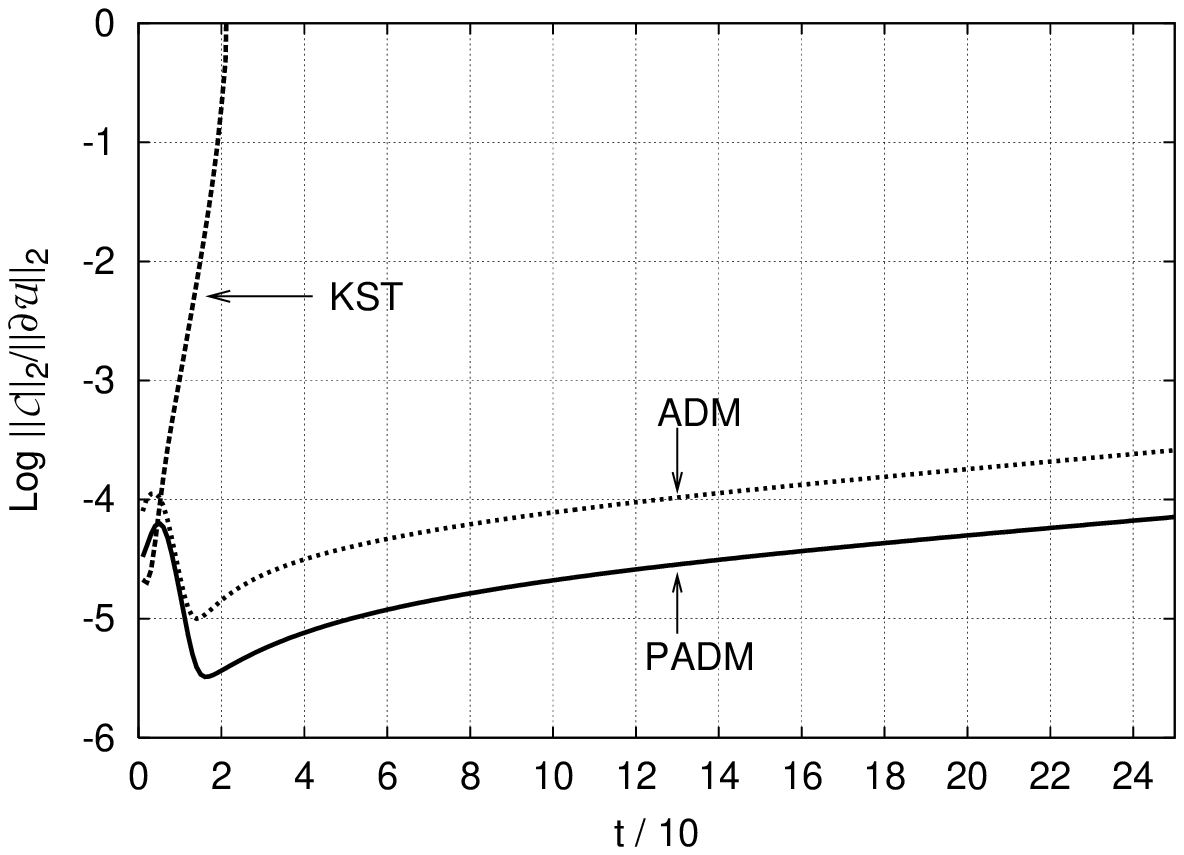}}
\subfigure[{\label{fig:4D}}]{\includegraphics[width=0.495\textwidth]{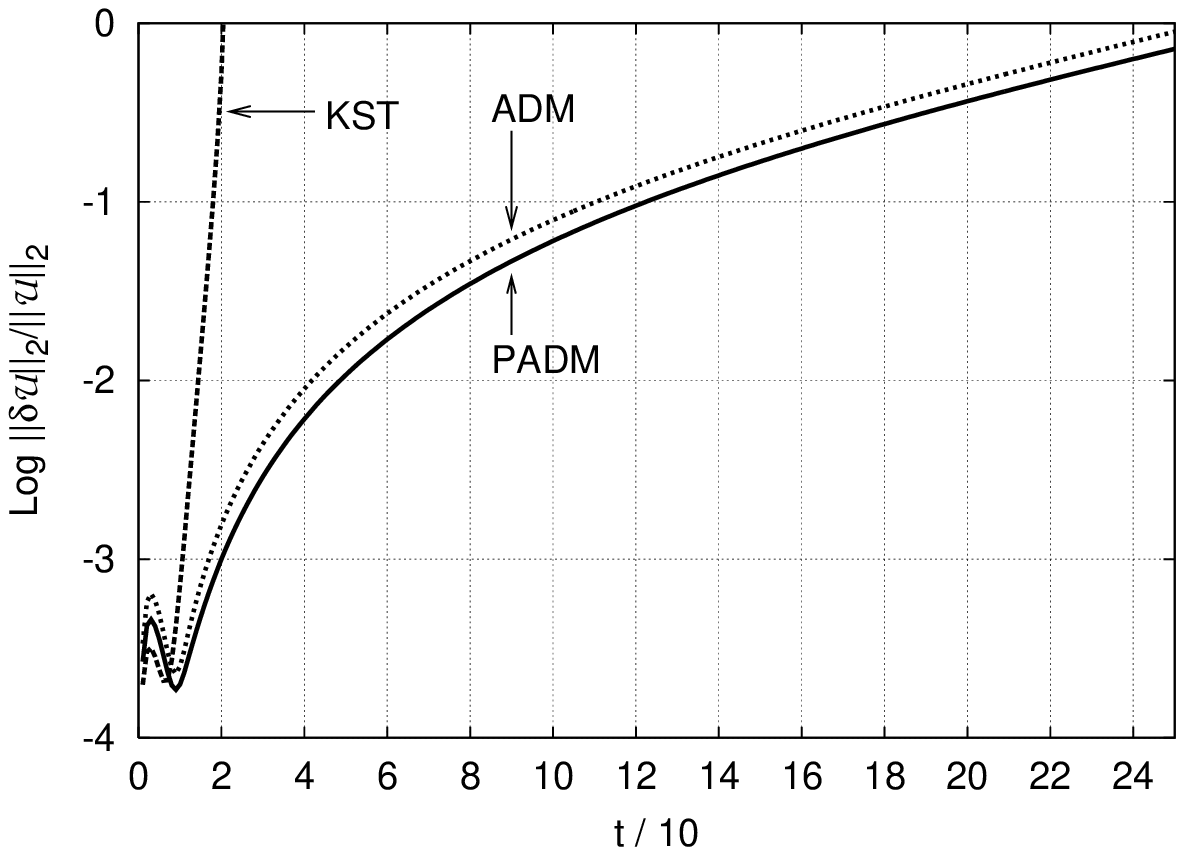}}
\caption{{Errors as functions of coordinate time $t$ for the expanding Gowdy wave for resolution $r=8$. On the left panel (a) we show the normalized constraint energies and on the right panel (b) we show the normalized  error energies for the ADM, KST and PADM formulations. \label{fig:4CD} }}
\end{figure*}

The results of the simulations are summarized in Figures \ref{fig:4CD}, where we plot the constraint energy (Fig. \ref{fig:4C}) and error energy (Fig. \ref{fig:4D})  for the ADM, KST and PADM systems for resolution $r=8$. These plots show that the runs with the KST formulation explode very quickly, whereas the ADM and PADM formulations can ran the test for significantly longer times.  The error energy and constraint energy of the PADM system lies lower than that of the ADM system and for this reason the PADM formulation extends the lifetime of the simulation by  12 light crossing times.

We checked that the simulations with the KST system do not crash because of violation of the numerical stability criterion. Instead, like in the gauge wave and robust tests, the error growth of the Gowdy simulations with KST system can be explained by the mixing and excitation of unwanted degrees of freedom the KST equations cause, because of  violations of the momentum constraints. However, the situation with the expanding polarized Gowdy wave is much more severe than the gauge wave. This is simple to understand by a glimpse at the evolution equation of ${\cal C}_{333}$ which is
\labeq{C333}{\begin{split}
\partial_t {\cal C}_{333}= & -\frac{3}{5}\alpha\gamma_{33} {\cal M}_3, 
\end{split}
}
Equation \eqref{C333} shows that, because of the exponential growth of both the lapse function and the three-metric with time, the ${\cal C}_{333}$ constraint is bound to grow exponentially with time. The RHS of \eqref{C333} varies roughly as $t^{-3}e^{\lambda/4}{\cal M}_3$. By using the analytic solution we can estimate the RHS of \eqref{C333}. For example, at $t=50$ even if ${\cal M}_3$ were ${\cal M}_3=10^{-16}$ the RHS of equation \eqref{C333} is of the order of $10^{46}$.

Although we are able to evolve expanding Gowdy wave initial data with the ADM and PADM systems to very late times, it is  seen in Figure \ref{fig:4D} that at the highest resolution the numerical solution matches the analytical solution up to approximately 110 light crossings with the ADM formulation and up to 122 light crossings with the PADM formulation. At these times the normalized error energy becomes of order $0.1$. The accuracy with the PADM system is completely lost after roughly $t=275$ and with the ADM system after roughly $t=258$. We note here that in \cite{KST_testbeds} the 1D (not 3D) simulations of the expanding gowdy wave with the KST system completely lose accuracy after 150 light crossing times, even though the authors use a much more sophisticated and accurate method of integration (i.e., pseudo spectral methods) in order to perform their simulations. For resolution $r=9$ we can accurately evolve the solution with the PADM system up to approximately 200 light crossings. 

Taking all these facts into accounts we conclude that the comparison of the three formulations in the expanding Gowdy wave testbed shows that the PADM formulation performs better than the ADM formulation and significantly better than the KST formulation.

\subsubsection{Collapsing Gowdy Cosmology}

\begin{figure*}
\subfigure[{\label{fig:4A}}]{\includegraphics[width=0.495\textwidth]{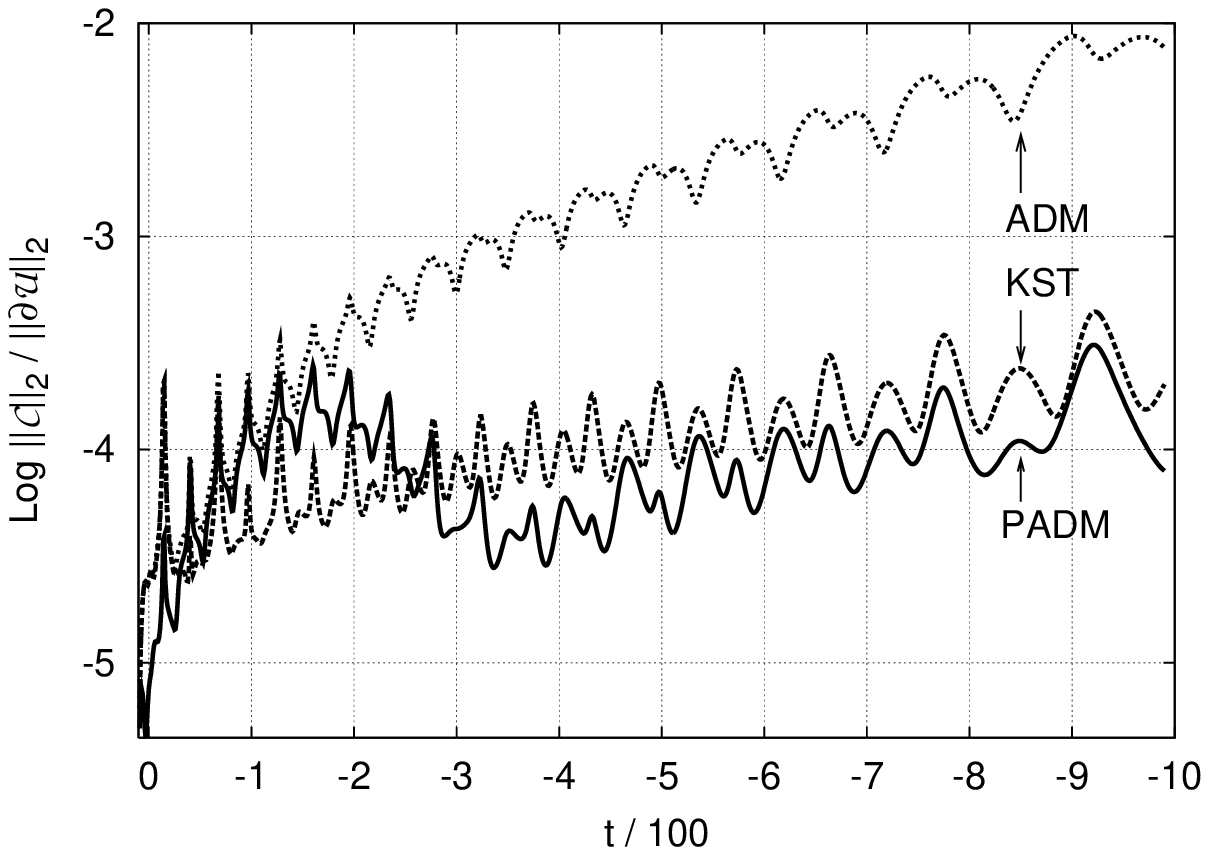}}
\subfigure[{\label{fig:4B}}]{\includegraphics[width=0.495\textwidth]{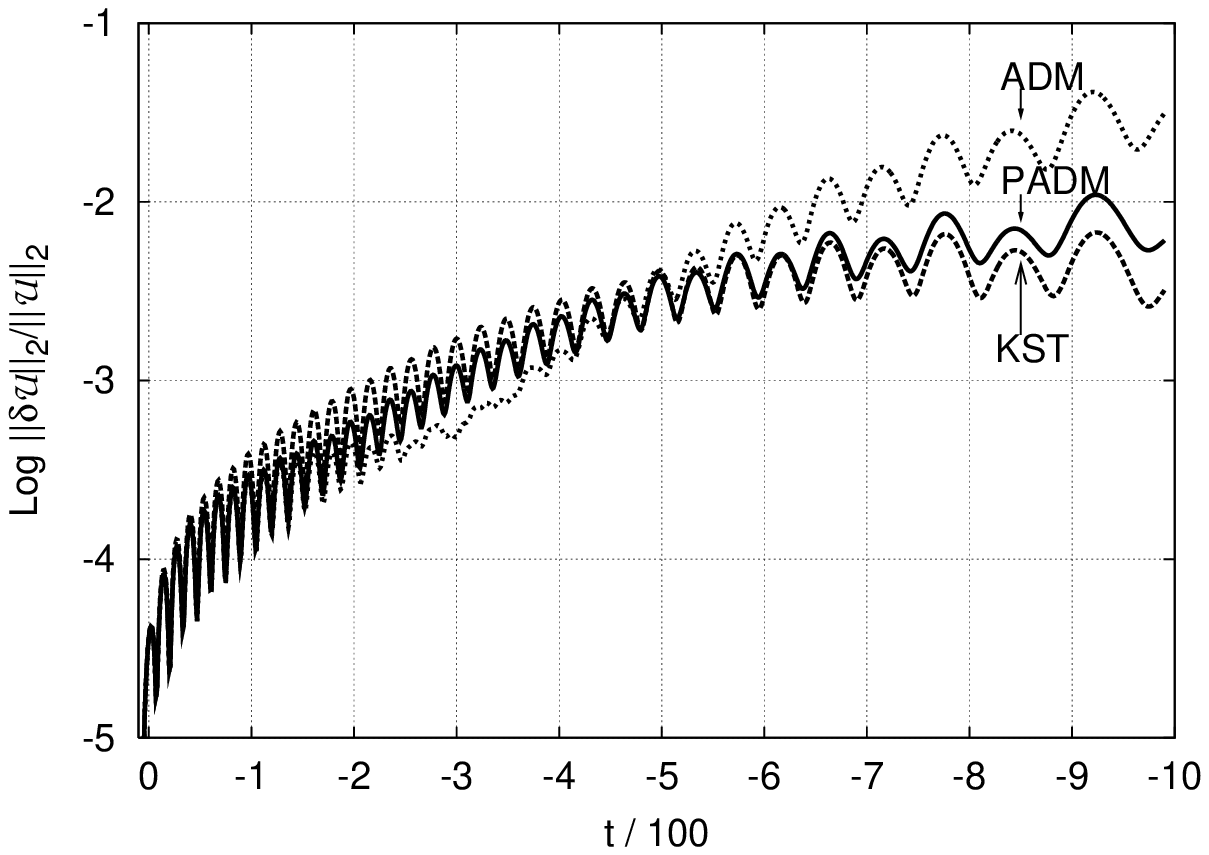}}
\caption{{Errors as functions of coordinate time $t$ for the collapsing Gowdy case for resolution r=7. On the left panel (a) we show the normalized constraint energies and on the right panel (b) we show the normalized  error energies for the ADM, KST and PADM formulations.   \label{fig:4.1} }}
\end{figure*}

 Unlike the expanding solution, which moves away from the $t=0$ singularity, the collapsing solution will reach the singularity very soon 
if integrated in the coordinate time $t$ as is dictated by the metric 
\eqref{gowdymetric1}. For this reason a coordinate transformation is implemented by asking for a new coordinate time $\tau$, such that $t=F(\tau)=ke^{c\tau}$ with $k,c$ constants given by $c \simeq 0.002119511921460752$ and $k\simeq 9.670769812764059$. In this new coordinate time the singularity is reached at $\tau=-\infty$ and the numerical integration is much more tractable. The metric in the new coordinates is given by
\labeq{coll_gowdy_metric}{
ds^2=-\alpha(\tau)^2 dt^2+F(\tau)^{-1/2}e^{\lambda(z,F(\tau))/2}dz^2+\tau dw^2,
}
where 
\labeq{lapse_coll_gowdy}{
\alpha(\tau)=cF(\tau)^{3/4}e^{\lambda(z,F(\tau))/4},
}
and
\labeq{coll_gowdy_metric2}{
dw^2=e^{P(z,F(\tau))}dx^2+ e^{-P(z,F(\tau))}dy^2.
}
The non-zero dynamical variables in the context of the first-order ADM formulation can be easily derived by use of \eqref{coll_gowdy_metric} and they 
are presented in appendix \ref{gowdy_coll_sol}.
The lapse function is harmonic and it is straightforward to show that 
\labeq{col_gowdy_lapse}{
\alpha=c\gamma^{1/2}.
}
We start the simulations at the 20th zero of the $J_0(2\pi t)$ Bessel function, i.e., at $\tau_o\simeq 9.87532058290983$ 
and evolve the initial data backwards in time using \eqref{col_gowdy_lapse}. We run the test for $r=6,7,8$ 
and for 1000 light crossing times with all formulations considered in this work. Finally, we note that for this test the values of the PADM parameters are opposite to those of equation \eqref{parabolic_params_val}.

The results of our simulations are summarized in Figures~\ref{fig:4.1} where we show the constraint energies and the error energies for all formulations considered here. We can see in \ref{fig:4A} that the ADM constraint energy grows in time and becomes larger than the PADM and KST constraint energies very early on. At $1000$ light crossing times the ADM constraint energy is about two orders of magnitude larger than the PADM and KST constraint energies. However, we see in Figure \ref{fig:4B} that the ADM error energy lies below the PADM and KST error energies for up to about $300$ light crossing times. After $500$ crossing times the ADM error grows faster than the PADM and KST error and at $1000$ crossing times there is roughly one order of magnitude difference between the ADM error energy and the corresponding error energies of the PADM and KST formulations.

Figures~\ref{fig:4.1} show that the PADM and KST systems perform equally well. However, for the most part of the 1000 crossing times the PADM constraint and error energies are less than those of the runs with the KST system. Towards the end of the simulation the KST error energy is slightly less than the PADM error energy, while the PADM constraint energy is slightly less than the KST constraint energy.


\section{Convergence \label{code_convergence}}


In this section we test the convergence and accuracy of our numerical schemes. In order to test the order of convergence we need to use exact solutions to the sets of PDEs which we solve. Of the four tests we performed for the comparison of the ADM, KST and PADM formulations, only the gauge wave and polarized Gowdy spacetimes are exact solutions of the Einstein equations and only these can form a real basis for convergence testing.

Unfortunately, the gauge wave results suffer by the excitation of exponential harmonic gauge modes and for this reason they cannot serve as a basis for convergence testing for long time. This is so, because of the properties of the solutions and not because of our numerical scheme. We demonstrated in section~\ref{gaugewave} that for different resolutions different exponential modes are excited and at different time-scales. Thus, convergence testing with the gauge wave can be done only for short periods of time before exponential modes become noticeable. We have been able to show that our code is indeed convergent and second-order accurate for short periods of time in the gauge wave test. 

The expanding Gowdy test cannot serve as a basis for convergence testing for long times either
because the test cannot be run for very long times. For example the KST formulation cannot even complete 22 light crossing times in the expanding Gowdy cosmology. Again this is because of the properties of the solution and not because of our numerical scheme. Nevertheless, we have been able to show that our code is convergent and second-order accurate in simulations of the expanding Gowdy wave with the ADM, KST, and PADM systems up until the simulations crash. In order to keep this discussion short, we omit the convergence plots for the expanding Gowdy and gauge wave tests, but they can be available by the authors upon request.

\begin{figure}
\includegraphics[width=0.35\textwidth,angle=-90]{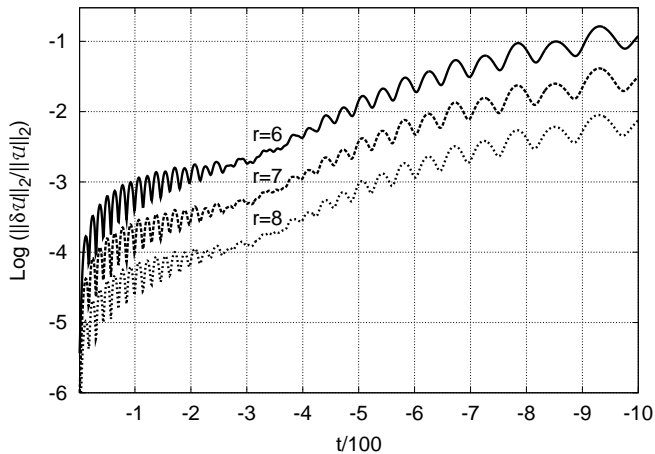}
\caption{Normalized error energies as functions of coordinate time $t$, for resolutions $r=6,7,8$. The plot corresponds to the simulations of the collapsing Gowdy spacetime with the ADM formulation.
 \label{fig:5A}}
\end{figure}

\begin{figure}
\includegraphics[width=0.35\textwidth,angle=-90]{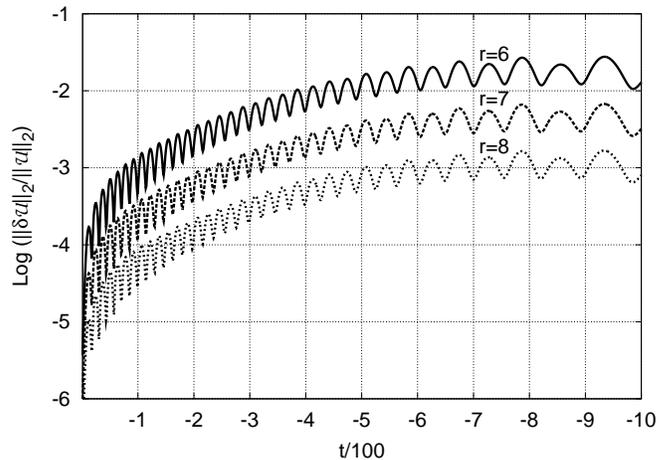}
\caption{Normalized error energies as functions of coordinate time $t$, for resolutions $r=6,7,8$. The plot corresponds to the simulations of the collapsing Gowdy spacetime with the KST formulation.
 \label{fig:5B}}
\end{figure}

\begin{figure}[h]
\includegraphics[width=0.35\textwidth,angle=-90]{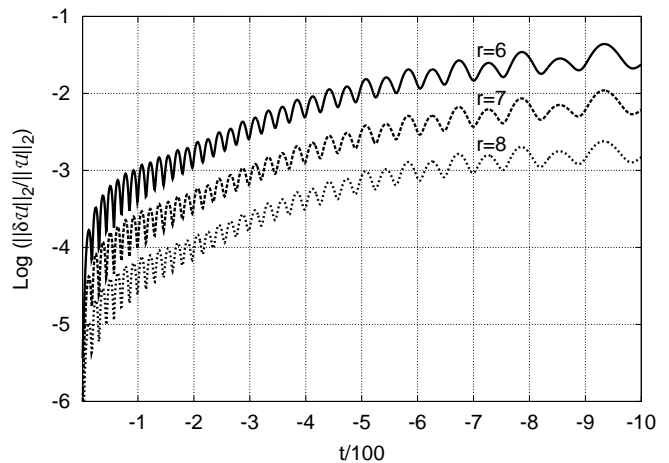}
\caption{Normalized error energies as functions of coordinate time $t$, for resolutions $r=6,7,8$. The plot corresponds to the simulations of the collapsing Gowdy spacetime with the PADM formulation.
 \label{fig:5C}}
\end{figure}

In order to test long term convergence of our code, we will use our results of the collapsing Gowdy wave testbed, where all three formulations successfully complete 1000 light crossing times. The convergence plots can be seen in Figures \ref{fig:5A}, \ref{fig:5B} and \ref{fig:5C} for the ADM, KST and PADM formulations respectively, where it is evident that with increasing resolution the numerical solution converges to the analytical one for all formulations.

\begin{center}
\begin{table} 
\begin{tabular}{cccc}\hline\hline
\multicolumn{1}{p{1.5 cm}}{\hspace{0.3 cm} $r_1, r_2$ } & 
\multicolumn{1}{p{1.5cm}}{ \hspace{0.35 cm}ADM \qquad} & 
\multicolumn{1}{p{1.5cm}}{\hspace{0.3 cm} KST }  & 
\multicolumn{1}{p{1.5cm}}{\hspace{0.15 cm} PADM \quad}  \\ \hline \hline 
$5, 6$ 		  & $1.9057$  		  &	$ 2.0461$          & $1.9617$                                                 \\  \hline
$6, 7$		         &  $1.9793$               &   $2.0146$          & $1,9945$                                               \\  \hline 
$7, 8$                &  $2.0296$                          &  $2.0032$                      & $2.0144$                                                 \\  \hline
\end{tabular}
\caption{Convergence order table. The first column shows the resolutions which are involved in the calculation of the order of convergence and the second to fourth columns show the order of convergence of our code with the ADM, KST and PADM formulations respectively.}
\label{table1}
\end{table}
\end{center}

It is also important to demonstrate that the order of convergence of our code is 2. To do this we use our data for the error energies and calculate the order of convergence (${\cal F}$), which is given by
\labeq{convergence}{
{\cal F}(t_n,r_1, r_2)= \frac{1}{r_2-r_1}\log_2\bigg(\frac{||\delta {\cal U}_N(t_n,r_1)||}{||\delta {\cal U}_N(t_n,r_2)||}\bigg),
}
where $r_1$ and $r_2$ two arbitrary resolutions,  $t_n$ a given time-step and $||\delta {\cal U}_N(t_n,r)||$ is the normalized error energy. Exact second-order accuracy means that
${\cal F}=2$.  In table \ref{table1} we show the time-average of ${\cal F}$ for the collapsing Gowdy spacetime with all three formulations and resolutions $\{r_1,r_2\}=\{5,6\}$, $\{r_1,r_2\}=\{6,7\}$ and $\{r_1,r_2\}=\{7,8\}$. The time-average order of convergence, $\overline{{\cal F}}$, is 
\labeq{}{
\overline{{\cal F}}(r_1, r_2)=\frac{1}{N}\sum_{n=0}^N {\cal F}(t_n, r_1, r_2),
}
where in our case $t_n=n$ and $N=1001$. Table \ref{table1} and Figures \ref{fig:5A}-\ref{fig:5C} clearly demonstrate that our code is convergent and second-order accurate.

\section{Conclusions \label{conclusions}}


We have described a stable, convergent and second-order accurate numerical scheme for solving the equations of the recently proposed PADM formulation of GR, we have tested the accuracy and stability of the PADM system and compared it with the first-order ADM and KST evolution systems. 

The ADM system is generally ill-posed, but in conjunction with a densitized lapse or ``1+log" slicing the 1D ADM evolution  equations are strongly hyperbolic and can be compared to other well-posed systems. The KST evolution system in conjunction with either a densitized lapse or ``1+log" slicing is strongly hyperbolic. The well-posed PADM equations have the structure of a mixed hyperbolic -- second-order parabolic set of PDEs. PADM possesses the desirable property of constraint damping. Unlike all other systems used in numerical relativity with constraint damping, the PADM system damps the short wavelength constraint violating modes more efficiently than the long wavelength ones. 

The series of tests which we used to compare the ADM, KST and PADM systems is known as the "Apples with Apples" or the "Mexico City" tests. These tests are designed to probe both the linear and the non-linear regime of the Einstein equations and they are called the robust stability  test, the gauge wave stability test, the linear wave stability test and the Gowdy wave stability test. The first three can be considered as perturbations about flat spacetime, so they probe the weak field limit of the Einstein equations, whereas the last one probes the strong field regime.

We ran the robust stability test with a densitized lapse with $\sigma=0.5,1$ and ``1+log" slicing  and we demonstrated that the PADM formulation performs much better than both the ADM and KST systems. The constraint violations with PADM are about seven orders of magnitude smaller than the constraint violations with KST system. PADM damps the constraints very quickly and the shorter the wavelength of the constraint violations the faster they are damped. This is in complete agreement with the  theoretical properties of PADM.  Finally, the ADM formulation crashes very quickly and faster for higher resolutions. Finally we demonstrated that the behavior of all three systems in this test is similar in the other gauge conditions we employed. This confirms the theoretical prediction that the PADM formulation damps the constraints independently of the gauge chosen.

The simulations of the high amplitude 1D gauge wave test clearly demonstrate that the PADM formulation performs better than the ADM formulation, while the KST system cannot evolve the initial data for more than 200 crossing times. The reason for this behavior of the KST system is that the dynamics of the formulation is such that it excites unwanted degrees of freedom, due to error in the momentum constraints. These degrees of freedom are not part of the dynamics of the gauge wave testbed and after they are excited, they grow exponentially with time and lead the computations to an end. 

The simulations of the high amplitude 2D gauge wave test also demonstrate that the numerical performance of the PADM formulation is much better than those of the ADM and KST systems. In this case it is the ADM simulations which crash very quickly. The runs with the KST equations  cannot evolve the initial data for more than 30 crossing times no matter how large the resolution is. Finally, the runs with PADM can successfully complete 100 light crossing times and show that the PADM system damps the constraints with time.

The linear wave testbed presented no essential difficulty for any of the three formulations, but both the 1D and the 2D linear wave simulations, show that the PADM formulation performs better than the ADM and KST formulations. In the 1D case the constraint error with the PADM system is quickly damped and is three orders of magnitude less than the constraint error with the ADM and KST systems. We observed similar behavior in the 2D linear wave, but the difference in the constraint error among the three formulations is not as large.

The polarized Gowdy wave testbed has two versions: the expanding one and the collapsing one. In the collapsing Gowdy wave, the KST and PADM systems perform equally well, but much better than the ADM one. At the end of the simulations the ADM constraint error is at least two orders of magnitude larger than the constraint error of the PADM and KST systems and the error in the solution with the ADM system is one order of magnitude larger than the corresponding errors with the KST and PADM systems. 

The expanding Gowdy wave was by far the strongest field solution we simulated. The crash of simulations of the expanding Gowdy solution, is naturally expected for any evolution system because the metric grows exponentially with time. The results of this test show that the PADM formulation performs better than the ADM formulation and dramatically better than the KST formulation. The expanding Gowdy wave initial data cannot be evolved with the KST system for more than 25 crossing times at the highest resolution we attempted. The cause for the sudden break down of the KST system is the same as the cause  for the sudden break down of the KST system in the 1D gauge wave. In contrast, the ADM and PADM systems can evolve the expanding Gowdy wave solution for significantly longer times, until the error in the solution grows so large that simulations terminate. The high resolution ADM runs completely lose accuracy after 258 light crossing times. The PADM runs are more accurate than the ADM ones and at the highest resolution completely lose accuracy after 275 light crossing times.

Taking all these facts into accounts, we conclude that PADM successfully passes the standard tests of numerical relativity and works equally well with a variety of algebraic gauges. Via the comparison of the numerical performance of the PADM formulation and those of the ADM and the KST formulations, we conclude that  PADM has better control of the constraint violations than both ADM and KST. The PADM system performs better than the ADM one and in most tests better than the KST system.

\appendix


\section{Expanding Gowdy Solution \label{gowdy_exp_sol}}


In this appendix we present the non-zero dynamical variables, in the context of the first-order ADM formulation, for the expanding Gowdy solution. Those can be  directly derived by use of equations \eqref{gowdymetric1}-\eqref{P_zt}, \eqref{ADM-gamma} and \eqref{D-defin} and they are given by
\labeq {}{
\gamma_{xx}=t e^P, \quad \gamma_{yy}=t e^{-P}, \quad \gamma_{zz}=t^{-1/2} e^{\lambda/2},
}
\labeq{}{\begin{split}
K_{xx} = & -\frac{1}{2\alpha}g_{xx}(t^{-1}+ \partial_t P), \\
K_{yy} = & -\frac{1}{2\alpha}g_{yy}(t^{-1}- \partial_t P), \\ 
K_{zz }= & \ \frac{1}{4\alpha}g_{zz}(t^{-1}-\partial_t \lambda),
\end{split}
}
where 
\labeq{}{
\alpha=\sqrt{\gamma_{zz}},
}
\labeq{}{\begin{split}
\partial_t\lambda(z,t)=& \
2 \pi ^2 t \big[(\cos (4 \pi  z)+1) J_1(2 \pi  t)^2 \\
 & \qquad \qquad +2 J_0(2 \pi  t)^2 \sin
   ^2(2 \pi  z)\big]
\end{split}
}
and
\labeq{}{
\partial_tP(z,t)=-2 \pi  J_1(2 \pi t) \cos (2 \pi  z).
}
We also have
\labeq{}{
D_{zxx}=\gamma_{xx}\partial_z P, \quad D_{zyy}=-\gamma_{yy} \partial_z P, \quad D_{zzz}=\half\gamma_{zz}\partial_z \lambda,
}
where 
\labeq{}{\begin{split}
\partial_z P(z,t)= & -2\pi J_0(2\pi t)\sin(2\pi z), \\
\partial_z\lambda(z,t) = & \ 4\pi^2t J_0(2\pi t) J_1(2\pi t)\sin(4\pi z).
\end{split}
}
\\


\section{Collapsing Gowdy Solution \label{gowdy_coll_sol}}


In this appendix we present the non-zero dynamical variables, in the context of the first-order ADM formulation, for the collapsing Gowdy solution. Those can be directly derived by use of equations  \eqref{coll_gowdy_metric}-\eqref{coll_gowdy_metric2}, \eqref{ADM-gamma} and \eqref{D-defin} and they are given by
\labeq{}{\begin{split}
\gamma_{xx}= & \ F(\tau) e^{P(z,F(\tau))}, \\
\gamma_{yy}= & \ F(\tau)e^{-P(z,F(\tau))}, \\ 
\gamma_{zz}= & \ F(\tau)^{-1/2}e^{\lambda(z,F(\tau))/2},
\end{split}
}
and
\labeq{}{\begin{split}
K_{xx}= & -\frac{c}{2\alpha}\gamma_{xx}(1+F\frac{\partial P}{\partial F}), \\ 
K_{yy}= & -\frac{c}{2\alpha}\gamma_{yy}(1-F\frac{\partial P}{\partial F}), \\
K_{zz}= & \ \frac{c}{4\alpha}\gamma_{zz}(1-F\frac{\partial \lambda}{\partial F}),
\end{split}
}
\\
where the lapse function $\alpha$ is given by \eqref{lapse_coll_gowdy}.
In addition, we have
\labeq{}{ \begin{split}
\pd{\lambda(z,F)}{F}= & \
2 \pi ^2 F \big[(\cos (4 \pi  z)+1) J_1(2 \pi  F)^2 \\
	& \qquad \qquad +2 J_0(2 \pi  F)^2 \sin^2(2 \pi  z)\big],
\end{split}
}
and
\labeq{}{
\pd{P(z,F)}{F}=-2 \pi  J_1(2 \pi F) \cos (2 \pi  z).
}

Finally the non-zero components of the $D_{kij}$ variables are given by
\labeq{}{
D_{zxx}=\gamma_{xx}\partial_z P, \quad D_{zyy}=-\gamma_{yy}\partial_z P, \quad D_{zzz}=\half\gamma_{zz}\partial_z \lambda, 
}
where 
\labeq{}{ \begin{split}
\partial_z P(z,F)= & -2\pi J_0(2\pi F)\sin(2\pi z),  \\
\partial_z\lambda(z,F)= & \ 4\pi^2F J_0(2\pi F) J_1(2\pi F)\sin(4\pi z).
\end{split}}
\\


\end{document}